\documentclass[letterpaper,twocolumn,10pt]{article}
\usepackage{usenix}

\usepackage{url}
\usepackage[utf8]{inputenc}
\usepackage{amssymb}
\usepackage{todonotes}
\usepackage{tabularx}
\usepackage{pifont}
\usepackage{algorithm}
\usepackage{algpseudocode}
\usepackage{listings}
\usepackage{amsthm}
\usepackage{caption}
\usepackage{subcaption}
\usepackage{array}
\usepackage{todonotes}
\usepackage{stmaryrd}
\usepackage{tikzsymbols}
\usepackage{breakurl}
\usepackage{enumitem}
\usepackage{float}
\usepackage{tcolorbox}
\usepackage{amsmath}
\usepackage{colortbl}
\usepackage{placeins}
\usepackage{dblfloatfix}
\usepackage{lineno}

\usepackage[available,functional,reproduced]{usenixbadges}

\setlist[itemize]{noitemsep, topsep=0pt, leftmargin=*}

\newtheorem{proposition}{Proposition}
\newtheorem{definition}{Definition}

\colorlet{mygreen}{green!50!black}
\newcommand{\myparagraph}[1]{\medskip\noindent\textbf{#1.}}
\newcommand{\myfirstparagraph}[1]{\noindent\textbf{#1.}}
\newcommand{\myanonparagraph}{\medskip\noindent}
\newcommand{\tick}[0]{\color{mygreen}\ding{51}\color{black}}
\newcommand{\cross}[0]{\color{red}\ding{55}\color{black}}
\newcommand{\meh}{\color{orange}$\mathbf{\sim}$\color{black}}
\newcommand{\concl}[1]{
    \begin{tcolorbox}[colback=lightgray!30, colframe=white, sharp corners, boxrule = 1pt, boxsep=3pt, left=0pt,right=0pt,top=0pt,bottom=0pt]
        \noindent{#1}
    \end{tcolorbox}
}

\DeclareCaptionLabelSeparator{dot}{. }

\newcolumntype{H}{>{\setbox0=\hbox\bgroup}c<{\egroup}@{}}
\newcolumntype{C}{>{\centering\arraybackslash}X}

\begin{document}

\title{Attacker Control and Bug Prioritization}

\author{
    {\rm Guilhem Lacombe}\\
   \textit{Université Paris-Saclay}, \textit{CEA}, \textit{List}, \textit{France}\\
   guilhem.lacombe.97@gmail.com
   \and
   {\rm Sébastien Bardin}\\
   \textit{Université Paris-Saclay}, \textit{CEA}, \textit{List}, \textit{France}\\
   sebastien.bardin@cea.fr
}

\date{}

\maketitle

\begin{abstract}
    As bug-finding methods improve, bug-fixing capabilities are exceeded, resulting in an accumulation of potential vulnerabilities.
    There is thus a need for efficient and precise bug prioritization based on exploitability. 
    In this work, we explore the notion of control of an attacker over a vulnerability's parameters, which is an often overlooked factor of exploitability.
    We show that taint as well as straightforward qualitative and quantitative notions of control are not enough to effectively differentiate vulnerabilities. 
    Instead, we propose to focus analysis on feasible value sets, which we call \textit{domains of control}, in order to better take into account threat models and expert insight. 
    Our new \textit{Shrink and Split} algorithm efficiently extracts domains of control from path constraints obtained with symbolic execution and renders them in an easily processed, human-readable form.
    This in turn allows to automatically compute more complex control metrics, such as \textit{weighted Quantitative Control}, which factors in the varying threat levels of different values. 
    Experiments show that our method is both efficient and precise.
    In particular, it is the only one able to distinguish between vulnerabilities such as \textit{cve-2019-14192} and \textit{cve-2022-30552}, while revealing a mistake in the human evaluation of \textit{cve-2022-30790}.
    The high degree of automation of our tool also brings us closer to a fully-automated evaluation pipeline.
\end{abstract}

\section{Introduction}

Over the past decades, program analysis research has predominantly focused on expanding and refining bug-finding methods.
As a result, automated techniques such as fuzzing, symbolic execution and data-flow analysis have improved significantly.
Fuzzing in particular is very prolific nowadays, with efforts such as Syzbot \cite{syzbot, syzkaller} uncovering thousands of bugs.
In addition, the popularization of open-source development models led to even greater bug-finding capabilities as third parties are able to contribute additional man- and processing-power.

\myparagraph{The problem} Unfortunately, bug-fixing capabilities have not seen the same level of development and thus unfixed bugs tend to accumulate over time.
{
This tendency is especially striking when looking at Syzbot's bug reports for the Linux kernel: 1561 open bugs as of December 2024, the oldest being over seven years old.
Furthermore, there are $51$ mentions of "out-of-bounds" and $120$ of "use-after-free".
Fuzzing of smaller scale projects as performed by Google's OSS-fuzz \cite{ossfuzz} also discovers large amounts of bugs, with over $1500$ labeled "vulnerability" in 2024 alone, including 427 buffer overflows, all of uncertain security impact.

This poses a severe security risk as overwhelmed developers are unable to identify true vulnerabilities and thus dispatch adequate bug-fixing efforts.}
Thus a need for precise yet efficient bug prioritization arises.

Due to the diversity of requirements and challenges inherent to different development projects, the choice of criteria for bug prioritization is ultimately up to the developers.
For example, bugs causing infinite loops may be of higher priority in a server than in an offline application due to the greater focus on availability.
However this choice is limited by the capabilities of current analyzers and error detectors, which typically focus on detecting \textit{classes} of vulnerabilities, such as use-after-free and out-of-bounds memory accesses in the case of the Linux kernel sanitizer KASAN \cite{kasan}.
For this reason, current bug prioritization practices often amount to simply attributing varying threat levels to different classes of vulnerabilities and are thus very coarse-grained.

In order to establish finer-grained prioritization criteria, one may consider using Automated Exploit Generation (AEG) \cite{Heelan, aeg} to automatically check for exploitability.
Ignoring the technical challenges that would be involved, there are still major issues with this approach, such as the fact that AEG gives no guarantees if no exploits are found.
Indeed, a lack of generated exploit only indicates incompatibility with the AEG method.
There is also a fundamental divergence in objectives: the goal of AEG is first and above all to build functional exploits,  not necessarily reusing the same exact bug given as input. 
Most AEG engines only use input bugs as starting points, some even only use them as seeds in their own bug search \cite{fuze, koobe, revery}. 

On the other hand, there are lessons to be learned from the inner workings or AEG engines.
In particular, many have some variant of a rating system for vulnerabilities in order to select the most adequate ones for the exploit being built.
Typically, vulnerabilities offering some level of control are preferred.
For example, KOOBE \cite{koobe} {compares} out-of-bounds write vulnerabilities based on obtainable offsets, sizes and written values, and prioritizes those with a wider range of capabilities.
{However its approach only allows for very limited partial ordering and is thus inadequate for generic bug prioritization.}
{\it An attacker's control over the value of vulnerability parameters is thus an important factor of exploitability.}

\myparagraph{Our goal}
We aim to develop fine-grained bug prioritization methods based on automatically estimating \textit{attacker control} over vulnerability parameters.

While attacker control over a variable or expression is generally understood to refer to the ability of an attacker to influence its value through program inputs, the lack of formal definitions renders this concept ambiguous.
Taint and symbolic bytes are often implicitly considered as qualitative indicators of control during taint analysis and symbolic execution respectively \cite{dtafse, Heelan, fuze}, despite a lack of formal guarantees.
On the other hand, Newsome et al. \cite{quantinfl} define a quantitative notion of control related to channel capacity, which assumes all values are equally dangerous and can thus be misleading.
For example, very large write sizes in a buffer overflow write will likely result in a crash, while smaller ones are more likely to hit critical data such as pointer or return addresses.
{Furthermore, applicable algorithms scale poorly and the authors did not explore applications to bug prioritization.}
\textit{Overall, attacker control lacks a unified theoretical framework as well as reasonably scalable approaches able to capture nuances such as the varying threat levels of different values.}

\myparagraph{Our proposal}
We propose to focus control analysis on the \textit{domains of control}, i.e., the sets of feasible values for vulnerability parameters.
Our approach consists in deriving the path constraint corresponding to program execution on a known vulnerability-triggering input using single-path symbolic execution, then use our \textit{Shrink and Split} algorithm to measure domains of control for a variable or expression in an analysis-friendly form.
{In particular, the absence of path exploration eliminates the risk of path explosion, bringing our symbolic execution more in line with dynamic taint analysis \cite{dtafse} in terms of scalability.}
Finally, vulnerabilities can be scored with control metrics based on these domains of control, such as \textit{weighted Quantitative Control}, the sum of the threat levels of the different feasible values according to some weight function. 
For example, we can compute a score for out-of-bounds write vulnerabilities by combining the weighted quantitative control over the write offset, size and data.  
While this work mainly focuses on {vulnerabilities involving well defined data-flows}, in particular out-of-bound memory bugs, we discuss applicability to other vulnerabilities as well.

\myparagraph{Contributions} 
We claim the following contributions: 

\begin{itemize}

\item  
{We provide a generic, unified and analysis-method-agnostic theoretical framework for the notion of attacker control over the value of a given expression.}
We define the \textit{domain of control} as the set of obtainable values (Section \ref{sec:defsdoms}). 
Then we define \textit{weak control} as the ability of an attacker to change the value through inputs and \textit{strong control} as the existence of inputs leading to every possible value (Section \ref{sec:defsqual}). 
Finally, we define \textit{quantitative control} (Section \ref{sec:qc}) based on existing notions and expand it into \textit{weighted quantitative control} (Section \ref{sec:wqc}).  
We then argue that \textit{taint analysis} can only ensure the absence of weak control (Section \ref{sec:tabad}), strongly limiting its usefulness for control analysis;

\item  
We propose \textit{Shrink and Split} (Section \ref{sec:algsns}), a novel algorithm able to efficiently determine domains of control as a set of intervals with weak and strong control guarantees.
It extracts them from logical formulas representing path constraints, {which can be obtained with standard  techniques such as symbolic execution or bounded model checking}.
Shrink and Split is a refinement process entirely based on qualitative analysis and yields approximate results when interrupted.
Additional regularity constraints, such as fixed bits, can also be taken into account to improve precision.
We also provide algorithms for measuring weak, strong and quantitative control;

\item
We implemented Shrink and Split and our other algorithms into a new dynamic analysis tool with symbolic execution capabilities.
Measuring control with Shrink and Split is significantly more precise than using taint analysis.
{Our algorithm also performs significantly better than the closest one from the literature (Newsome et al. \cite{quantinfl}, see Section \ref{sec:xps}). }
Furthermore, Shrink and Split is more robust and more informative than counting algorithms.
Our tool also implements various utilities enabling automatic end-to-end analysis; 

\item
We perform a ground-truth evaluation on real-world vulnerabilities (Section \ref{sec:rq3}) which shows that our approach gives a more precise exploitability assessment than CVSS scores and quantitative information flow analysis. 
In particular, we are able to clearly differentiate vulnerability capabilities, match identical ones and even {help to} correct previous human evaluation in the case of \textit{cve-2022-30790}.
The high degree of automation of our tool also allowed us to evaluate memory out-of-bounds bugs from the Magma fuzzing benchmark \cite{magma} in an end-to-end manner and realistic field conditions (Section \ref{sec:rq4}).
We were additionally able to show that the results from this experiment match the tendencies of coarse-grained human expectations.

\end{itemize}

This work constitutes a step toward precise, fully-automated exploitability assessment.
While exploitability as a concept is difficult to formally grasp and may not be reducible to a single catch-all notion, control is definitively an interesting part of it. Combinations with other indicators is an interesting research direction.

\section{Motivating Example}

\label{sec:motex}

\begin{figure}[!htb]
\begin{minipage}{\columnwidth}
    \centering
    \begin{lstlisting}[
        language=c, 
        caption=Motivating example, 
        label=lst:motex,
        basicstyle=\scriptsize,
        commentstyle=\color{gray},
        numbers=left,
        xleftmargin=3em,
        xrightmargin=1em,
        frame=single,
        captionpos=b
    ]
#define HEADER_SIZE 40

uint64_t check_header(char *input, 
        uint64_t input_size)
{
    //2) input[0->7] written on the stack
    uint64_t header = *((uint64_t *) input);
    return header <= 296;
}
    
void get_msg(char *buf, uint64_t buf_size, 
        char *input, uint64_t input_size)
{
    //3) not initialized => size = header
    uint64_t size;
    if(input_size <= buf_size + HEADER_SIZE)
        //4) input_size < 40 => integer underflow
        size = input_size - HEADER_SIZE;
    //5) buffer overflow!!!
    // a. input_size < 40 
    //    => 2^64 - 40 <= size < 2^64
    // b. input_size > (*@{296}@*) => size = header 
    memcpy(buf, input + HEADER_SIZE, size);
}

int main(...)
{
    //1) inputs: char *input, uint64_t input_size
    ...
    char buf[256];
    if(check_header(input, input_size))
        get_msg(buf, 256, input, input_size);
    ...
}
    \end{lstlisting}
\end{minipage}
\end{figure}

In order to illustrate the problem of bug prioritization, consider the program from Listing \ref{lst:motex}, which exhibits two similar vulnerabilities with different levels of exploitability.

\myparagraph{Explanation} 
We assume that $input\_size$ and $input$ are independently provided by the attacker, thus $input\_size$ is not constrained by $input$.
However the program assumes that it is and that $input$ has a header of size $HEADER\_SIZE$ that must be removed.
This leads to unintended behaviour when passing malformed inputs.

\textbf{\textit{Vulnerability a}} is triggered when $input\_size$ is smaller than $HEADER\_SIZE$.
In this scenario, the branch condition on line $16$ is true, leading to an integer underflow on line $18$.
A buffer overflow then occurs at line $23$ with $2^{64} - 40 \leq size < 2^{64}$, the size of $buf$ being only $256$.

\textbf{\textit{Vulnerability b}} occurs when $input\_size$ is greater than $buf\_size + HEADER\_SIZE$.
This time, the branch condition on line $16$ is false and $size$ remains uninitialized.
Its value is therefore equal to the first $8$ bytes of $input$ which were written on the stack on line $7$.
The attacker can thus obtain $257 \leq size \leq 296$ on line $23$ and cause a buffer overflow.

\myparagraph{Discussion} 
While the main impact of both vulnerabilities is a buffer overflow allowing an attacker to overwrite the stack, their levels of exploitability differ greatly.
Indeed vulnerability $a$ always results in a crash as the program attempts to overwrite roughly $2^{64}$ bytes of memory.
On the other hand, vulnerability $b$ allows an attacker to overwrite at most $40$ bytes of memory with data they provide. 
This level of \textit{control} makes vulnerability $b$ easier to exploit.
For example, the attacker can overwrite the return address of the $main$ function and hijack control-flow.

In the context of security-aware bug prioritization, vulnerability $b$ should be fixed before vulnerability $a$.
However it would be difficult to recognize this fact in practice when only vulnerability types are reported.
In this case, information about how \textit{controllable} vulnerabilities $a$ and $b$ are is required to give them the correct level of priority.
Our experimental evaluation shows that this scenario can indeed happen in practice (see Table \ref{tab:scores} in Section \ref{sec:xps}).

\myparagraph{Prior work} Commonly used indicators of control such as taint and symbolic bytes \cite{dtafse, fuze} are not helpful here: $size$ would be tainted and symbolic on line $23$.
It also can take the same number of values in both vulnerabilities thus quantitative information flow analysis \cite{klebanov_qif} is equally unhelpful.
Correctly handling such cases requires a new approach, which we detail in Sections \ref{sec:def} and \ref{sec:algs}. 

\section{Background}

\label{sec:bg}

Let $P$ be a program.
An execution of $P$ is a sequence of states each associated with a location corresponding to the current instruction.
Let $S_P$ be the set of possible states of $P$, $L_P$ the set of locations of $P$ and $\lambda(s) \in L_P$ the program location associated with a state $s$.
We note $\rightarrow : S_P \times S_P$ the transition relation between states during execution of $P$ and $\rightarrow^*$ its transitive closure.

Sequences of states  are called \textit{traces} and sequences of locations \textit{paths}.
A path $\pi = (l_i)_i$ is said to be feasible \textit{iff} there exists a possible trace $t = (s_i)_i$ of $P$ following $\pi$, i.e., forall $i$, $\lambda(s_i) = l_i$. 
If $\pi$ starts with $s$ and ends with $s'$, we note $s \rightarrow_\pi s'$. 

Let $V_P$ be the set of variables of $P$ and $I_P$ its set of possible inputs.
We note $s(v)$ the value of a variable $v$ in a state $s$ and $Dom(v)$ the  set of values that $v$ can take a priori (e.g., considering its type).
To simplify notations, \textit{we assimilate expressions over variables as implicit variables.}
For any input $i \in I_P$ and state $s \in S_P$, we note $i \rightarrow^* s$ \textit{iff} $s_0 \rightarrow^* s$ with $s_0$ the initial state corresponding to input $i$.

\subsection{Taint Analysis} 

Taint analysis \cite{dtafse} refers to the static or dynamic propagation of tags associated to data to qualitatively track data-flows.
An expression introducing a tag is referred to as a \textit{source} while program locations where the presence of  tags is checked are called \textit{sinks}. 
This technique has the advantage of being fairly lightweight at the cost of limited precision and guarantees. 
A common application is the verification and enforcement of non-interference properties \cite{sec_pols_and_models}.
Some prior works use taint analysis to detect attacker control \cite{Heelan,ardilla}.
 
\subsection{Symbolic Execution}

Symbolic Execution (SE) \cite{dtafse} refers to the exploration of execution paths in a program while computing the associated constraints on inputs and other variables such as registers and memory locations.
These constraints usually take the form of a Satifiability Modulo Theory (SMT) formula and are used to check whether paths are feasible.
Variables dependant on inputs are referred to as \textit{symbolic} variables while constant values are said to be \textit{concrete}. 
While SE excels at covering edge-case paths and allows to generate inputs triggering them, it may suffer from \textit{path explosion} , i.e., an exponential increase in the number of paths being followed.
Some constraints may also be too difficult to solve (typically, cryptography).
SE engines include KLEE \cite{klee} at LLVM-IR level as well as Angr \cite{angr} and BINSEC \cite{binsec, binsec_se} for binary analysis.

\myparagraph{Choice of SMT theory}
We assume that symbolic states are expressed in the $ABV$ theory (arrays and bitvectors), with variables represented as bitvectors and memory as a logical array.
As a consequence, the values of variables are \textit{finite}.

\myparagraph{Notations}
We note $SE(\pi)$ the symbolic state obtained at the end of the symbolic execution of a path $\pi$, with program inputs initially symbolic.
Let $\phi$ be a symbolic state.
We define the following notations:
\begin{itemize}
    \item $\phi(x) \triangleq True$ if $x$ satisfies $\phi$, $False$ otherwise
    \item $sat(\phi) \triangleq True$ if $\exists$ $x: \phi(x)$, $False$ otherwise
    \item $val(\phi,v)$ a feasible value of $v$ in $\phi$, $nil$ if none exists
    \item $val(\phi(x),v)$ the value of $v$ for the input $x$ if $\phi(x)$, else $nil$
    \item $duplicate(\phi,v) \triangleq \phi', v'$ with $\phi'$ a copy of $\phi$ with separate variables, $v'$ being equivalent to $v$
\end{itemize}

\subsection{Quantitative Information Flow}

\label{sec:backqif}

Quantitative information flow analysis consists in evaluating the quantity of information flowing through a channel, often by measuring channel capacity \cite{chancapa}.
This is usually done to quantify the leakage of secret information \cite{clark, heusser, mccamant}.

\myparagraph{Projected Model Counting (PMC)} 
While model counting counts the number of solutions of a formula, PMC counts the  number of values a subset of free variables can take within a SAT formula \cite{pmc}. 
It can be used to measure QIF by converting SMT formulas generated with symbolic execution or model checking into SAT formulas \cite{klebanov_qif, approxflow}.
We note $SE_{PMC}$ the algorithm combining SE and PMC.

While the precision of current exact PMC solvers such as D4 \cite{d4} and Ganak \cite{ganak} is high, scalability is an issue and they may even fail on trivial examples as shown in Section \ref{sec:xps}.
ApproxMC \cite{approxmc_CMV16, approxmc_SM19, approxmc_cav20} is an efficient approximated PMC solver with probably approximately correct guarantees, i.e., results are within a given error margin with a given probability. 

\myparagraph{Quantitative Influence}
Newsome et al. \cite{quantinfl} defined the notion of quantitative influence, derived from channel capacity, as the logarithm of the number of feasible values for a variable.
We integrate this notion into our control framework as \textit{quantitative control} in Section \ref{sec:qc}. 
They also proposed an algorithm for measuring quantitative influence based on statistical sampling, which we discuss in Section \ref{sec:algsns} and evaluate against in Section \ref{sec:xps}.

\section{Problem Statement}

\label{sec:scope}

Our goal is to rank vulnerabilities according to how controllable their parameters are, with as little assumptions on the goal of a potential attacker as possible, and as little human effort as possible.

\subsection{Requirements}

\myparagraph{Vulnerability information}
We assume that the \textit{type of vulnerability} (e.g., out-of-bounds write, use-after-free, pointer corruption, etc.) is known.
We also assume that we have a \textit{triggering input}. 
These parameters can be expected to be available for most discovered vulnerabilities as proof of existence and for debugging purposes. 
Our method can also be paired with automated bug finding methods such as symbolic execution \cite{dtafse} or fuzzing \cite{fuzz} to directly find inputs. 

\myparagraph{Human expertise}
Identifying vulnerability parameters, which we refer to as \textit{target variables}, may require some human expertise, although they can be identified automatically in most cases as shown in Section \ref{sec:rq4}. 
These parameters vary depending on vulnerability types: size and data of a buffer overflow write, value of a corrupted pointer, etc.
They can be implicit, meaning that the value itself is not stored anywhere at runtime, such as the overall write size from a \textit{strcpy} call.

\subsection{Archetypal Control Problems}
Notions of control can relate to a wide array of vulnerabilities in many ways.
Rather than considering individual types of vulnerabilities, we define archetypal control problems:

\myparagraph{Data control problem}
This problem is the most straight forward: the vulnerability allows an attacker to control the value of a well-defined structure, such as a pointer or a used-after-free object.
In this case, we only need to analyze the degree of control of the attacker over the value of this structure.

This archetype fits vulnerabilities such as some use-after-frees, use-before-initialization and pointer corruption.

\myparagraph{Memory range control problem}
This problem covers vulnerabilities allowing an attacker to read or write variable size data at variable memory addresses, without any explicit structure.
Here the vulnerability's parameters are the address, size and written data when applicable.
As variable-size data can be difficult to handle, one may resort to extrapolating from fixed-size analysis.

This archetype fits most out-of-bounds read / write vulnerabilities, such as vulnerabilities $a$ and $b$ from Listing \ref{lst:motex}.

\paragraph{What about vulnerabilities without explicit data-flows?}
Control is inherently tied to data-flows.
However, the effect of those data-flows are not always materialized into an explicit value.
{For example, attackers may control the number of memory allocator interactions in order to achieve heap layout manipulation \cite{shrike,gollum,maze}.
In such cases, parameters must be made explicit at the code level or within the analysis, for example with counters.}

\myparagraph{Scope}
\textit{We choose to mostly focus on memory range problems in this work} (e.g., buffer overflows) to avoid scope creep {and since they match the typical bugs found by fuzzers.}
Our benchmark nevertheless also contains examples of the data problem (e.g., pointer corruption after a use-after-free).

\section{Formally Defining Control} 

\label{sec:def}

In order to explicitly define various aspects of the notion of control, we first need to precisely identify the meaning of \textit{controlling a variable} from a natural language standpoint.
We consider attackers interacting with a program through its inputs, with the goal to take advantage of a given vulnerability (e.g., buffer overflow) with several variable parameters (e.g., the size  of a buffer overflow write), for some malicious purpose (e.g., rewriting a return address or a function pointer). 

The \textit{weakest level of control} attackers are looking for is the ability to \textit{influence} the value of the vulnerability parameters through inputs, i.e., to obtain different values for them.
In other words, attackers are not satisfied with simply knowing of the existence of a data-flow from inputs to the targeted (parameter) variables, it must also be influenceable. 
On the other hand, the \textit{strongest level of control} the attacker is seeking is the ability to obtain any value for the vulnerability parameters.

We propose hereafter a hierarchy of notions of control with precise formal definitions. 
All will be relevant in Section \ref{sec:algs} when we propose algorithms for control evaluation.

\subsection{Domain of Control}

\label{sec:defsdoms}

First let us define our key notion of \textit{Domain of Control}, from which we derive all our other notions of control. 

Given a program $P$, we define the \textit{Domain of Control} of variable $v$ at location $l$ as the set of feasible values for $v$ at this location, i.e., the set of values $e$ of $v$ for which we can find some input $i$ such that executing $P$ on $i$ leads to a program state $s$ (at location $l$) in which $v$ evaluates to $e$. 
More formally, we have the following definition:

\begin{definition}[Domain of Control]
    \[DoC(v,l) \triangleq \{e \in Dom(v) / \exists i \in I_P: i \rightarrow^* s \in S_P\]
    \[\text{with } \lambda(s) = l \text{, } s(v) = e\}\]
\end{definition}

\subsection{Qualitative Control}

\label{sec:defsqual}

We now define \textit{Weak Control} -- resp.~\textit{Strong Control} -- as the attacker's ability to find inputs of the program leading to different values -- resp.~leading to any value -- of $v$ at $l$. 

\begin{definition}[Weak Control (WC)]
    Given a program $P$, a variable $v \in V_P$ is {weakly controlled} at location $l \in L_P$ if there exists $i, i' \in I_P$ two inputs such that $i \rightarrow^* s$ and $i' \rightarrow^* s'$ with $s, s' \in S_P$, $s(v) \neq s'(v)$ and $\lambda(s) = \lambda(s') = l$. 
    We note $WC(v,l)$.
\end{definition}

Note that $v$ is weakly controlled at $l$ \textit{iff} $|DoC(v,l)| > 1$, hence \textit{iff} $v$ can indeed take at least two values, depending on the input chosen by the attacker.

\begin{definition}[Strong Control (SC)]
    We say that $(v,l)$ is strongly controlled if for all $e \in Dom(v)$, there exists an input $i \in I_P$ such that $i \rightarrow^* s \in S_P$ with $s(v)=e$ and $\lambda(s)=l$.  
    We note $SC(v,l)$.
\end{definition}

In this case, $v$ is strongly controlled at $l$ \textit{iff}  $|DoC(v,l)| = |Dom(v)|$, hence \textit{iff} $v$ can take any value from its definition domain. 
Intuitively, $SC$ is a stronger property than $WC$.

\begin{proposition}
   Strong control is stronger that Weak control, i.e., for any program $P$ and target $(l,v)$,  $SC(v,l) \Rightarrow WC(v,l)$. 
\end{proposition}

\begin{figure}[!htb]
\begin{minipage}{\columnwidth}
    \centering
    \begin{lstlisting}[
        language=c, 
        caption=Examples of Weak and Strong Control, 
        label=lst:exwcsc,
        basicstyle=\scriptsize,
        commentstyle=\color{gray},
        numbers=left,
        xleftmargin=3em,
        xrightmargin=1em,
        frame=single,
        captionpos=b
    ]
int x = input;
//here x is strongly controlled
if(x)
    //here x is weakly but not strongly controlled
if(!x)
    //here x is neither weakly nor strongly 
    //controlled
    \end{lstlisting}
\end{minipage}
\end{figure}

The main drawback of these definitions is a lack of nuance, as weak and strong control guarantee the lowest and highest amount of control possible respectively as shown on Listing \ref{lst:exwcsc}.
In the case of our motivating example (Listing \ref{lst:motex}), the out-of-bounds write size is weakly but not strongly controlled for both vulnerabilities, thus these notions cannot be used to distinguish between them.
Nevertheless, we show in Section \ref{sec:algsns} that $WC$ and $SC$ can be used as formal guarantees within more subtle approaches.

\subsection{Quantitative Control}

\label{sec:qc}

Shifting to quantitative measurement is a common approach when attempting to overcome the limits of qualitative analysis \cite{heusser, azizVerifyingContinuousTime1996, hanssonLogicReasoningTime1994, probse}.
Following this trend and similarly to Newsome et al. \cite{quantinfl}, we define \textit{Quantitative Control} as the (normalized) channel capacity \cite{chancapa} of $v$.

\begin{definition}[Quantitative Control (QC)]
    (similar to Newsome et al. \cite{quantinfl})
    \[QC(v,l) \triangleq \frac{ln(|DoC(v,l)|)}{ln(|Dom(v)|)}\]
\end{definition}

The value of $QC$ relates intuitively to $WC$ and $SC$.

\begin{proposition}
    $WC(v,l)$ $\iff$ $QC(v,l) > 0$
\end{proposition}

\begin{proposition}
    $SC(v,l)$ $\iff$ $QC(v,l) = 1$
\end{proposition}

While $QC$ allows for a level of precision well beyond $WC$ and $SC$, measuring it in practice is challenging due to the poor scalability of counting algorithms (see Section \ref{sec:xps}).
In addition, $QC$ remains a one-dimensional measure of control: it assumes that all possible values are equally dangerous.
However, this is often not the case as in our motivating example (Listing \ref{lst:motex}), where both vulnerabilities have the same number of write sizes but the larger write sizes of \textit{vulnerability a} lead to a crash and thus have a lower threat level.

\subsection{Weighted Quantitative Control}

\label{sec:wqc}

In light of the limitations of $WC$, $SC$ and $QC$, we argue that fine-grained evaluation of attacker control should focus on the domains of control and account for expert insight into which values are more dangerous. 

For instance, in our motivating example (Listing \ref{lst:motex}), we see that buffer overflow write sizes for \textit{vulnerability a} are too large to be useful as they will lead to crashes, while those for \textit{vulnerability b} allow to overwrite nearby stack data only and are therefore more useful for the attacker. 
Generally, smaller overflow write sizes / offsets should be worth more due to the lower variability of near targets and their potential interest for the attacker (chunk metadata, adjacent objects, return addresses and other stack data, etc.).

One way to automatize such reasoning is to compute a weighted quantitative control metric. 

\begin{definition}[Weighted Quantitative Control (wQC)]
    Let $\omega_v : Dom(v) \rightarrow \mathbb{R}$ be a weight function.
    \[wQC(v,l,\omega) \triangleq \frac{\sum_{n \in DoC(v,l)} \omega(n)}{\sum_{n \in Dom(v)} \omega(n)}\]
\end{definition}

This may be difficult to compute in practice, however in some instances it is possible to efficiently approximate.

\begin{proposition}
    \label{prop:wqcint}
    Let $Dom(v) = \llbracket a, b\rrbracket \subset \mathbb{N}$, $P$ be a set of intervals in $\mathbb{N}$ and a partition of $DoC(v, l)$.
    Let $\Omega : \mathbb{R} \rightarrow \mathbb{R}$ be an integrable function on $[a, b]$ and $\omega = \Omega|_{Dom(v)}$.
    \begin{equation}
        \label{eq:wqcint}
        wQC(v, l, \omega) \approx \frac{\sum_{\llbracket i, j\rrbracket \in P} \int_i^{j + 1} \Omega(x)dx}{\int_a^{b + 1} \Omega(x)dx}
    \end{equation}

    If we have additional constraints on intervals\footnote{For example, $x \equiv 0 [2]$}, with $\rho(I)$ the actual number of feasible values in $I \in P$, we get:
    \begin{equation}
        \label{eq:wqcconstr}
        wQC(v, l, \omega) \approx \frac{\sum_{\llbracket i, j\rrbracket \in P} \frac{\rho(\llbracket i, j \rrbracket)}{j - i} \int_i^{j + 1} \Omega(x)dx}{\int_a^{b + 1} \Omega(x)dx}
    \end{equation}
\end{proposition}

For example, for buffer overflow / underflow write vulnerabilities such as in our motivating example, we can use $\omega : x \mapsto \frac{1}{ln(2)x}$ for out-of-bounds write sizes and offsets, in order to introduce bias toward smaller values, i.e., tampering of local data.
This way, the weighted quantitative control value for the overflow size, i.e., the write size minus the size of the buffer, is approximately $\frac{\int_{2^{64}-296}^{2^{64}-256} \omega(x)dx}{\int_{1}^{2^{64}} \omega(x)dx} = \frac{log_2(2^{64}-256) - log_2(2^{64} - 296)}{log_2(2^{64})} \approx 0$ for \textit{vulnerability a} and $\frac{\int_1^{41} \omega(x)dx}{\int_{1}^{2^{64}} \omega(x)dx} = \frac{log_2(41)}{log_2(2^{64})} \approx 0.0837$ for \textit{vulnerability b}, thus giving a clear order of priority.

\smallskip 

Exact, constraint-based domain of control representations, such as SMT formulas, are often too complex to perform such analysis.
We present a solution to provide a simplified representation of $DoC(v,l)$ 
in Section \ref{sec:algs}.   

\subsection{Variants} 

\myfirstparagraph{Control under assumption}
So far, we have defined control over the entirety of $Dom(v)$.
However it can be useful to verify control properties over a subset $E \subset Dom(v)$ of feasible values, as it would allow to factor in assumptions such as $v$ being always even or within a given range.
We thus define notions of control over a subset of values similarly to their regular counterparts with $Dom(v)$ reduced to $E$, denoted with the suffix $|_E$.
In practice, we make use of the notion of strong control under assumption in Algorithm \ref{alg:sas} in Section \ref{sec:algsns}.

\myparagraph{Restriction to a single path}
Reasoning over a single path is advantageous in the context of program analysis as it eliminates loops. 
We define $DoC_\pi$, $WC_\pi$, $SC_\pi$, $QC_\pi$ and $wQC_\pi$ similarly to their non-single-path counterpart, with the constraint that values are obtainable through a single path $\pi$, i.e., $\rightarrow^*$ is replaced with $\rightarrow_\pi$.

While this approach comes at the cost of completeness, many vulnerabilities can still be adequately analyzed this way. 
Furthermore, domains of control for different paths can be merged without loss of precision.
The limits of this approach and potential solutions are further discussed in Section \ref{sec:disc}.

\subsection{Conclusion}

To sum up, weak and strong control are too extreme to be insightful on their own. Quantitative control allows for more precision, however it still remains uni-dimensional.
This is an issue since control is a complex, multi-dimensional notion with different values potentially having different threat levels.
We propose to focus more on this aspect by directly analyzing domains of control and distilling them down with more expressive metrics, such as weighted quantitative control.

\section{Evaluating Control} 

\label{sec:algs}

Assuming vulnerability-triggering inputs are available, we choose to employ dynamic analysis methods at binary level in order to precisely follow complex program behaviour at the cost of restricting analysis to a single execution path $\pi$.

We will first discuss why taint is a poor indicator of control, before detailing symbolic-execution-based solutions.

\subsection{Taint Analysis Cannot Guarantee Control}

\label{sec:tabad}

A typical taint policy consists in propagating taint to the outputs of an instruction if at least one of its inputs is tainted (see Appendix \ref{fig:appdta} for an example).
While one may be tempted to interpret taint as an indicator of control, it unfortunately cannot give much guarantees.

\begin{proposition}[Taint is limited for control evaluation]
    Taint only guarantees that $\lnot tainted \Rightarrow \lnot WC$. 
\end{proposition}

\begin{figure}[!htb]
\begin{minipage}{\columnwidth}
    \centering
    \begin{lstlisting}[
        language=c, 
        caption=Weak Control false positives with taint analysis, 
        label=lst:extaint,
        basicstyle=\scriptsize,
%
        commentstyle=,
%
        numbers=left,
        xleftmargin=3em,
        xrightmargin=1em,
        frame=single,
        captionpos=b
    ]
int x = input; (*@$\leftarrow$ \textit{tainted}@*)
int y = input;
int z = x + y;
if(x >= 0)
{
    if(x <= 0) (*@$\leftarrow$ \textit{x tainted but can only be equal to 0}@*)
}
int w = z - x; (*@$\leftarrow$ \textit{w tainted but can only be equal to y}@*)
    \end{lstlisting}
\end{minipage}
\end{figure}

If a variable is tainted, there is no guarantee that its value can change depending on inputs, i.e., that it is weakly controlled.
{While taint can be removed when a single instruction restricts a variable to a single value (e.g., $if(!x)$), this can also happen due to arbitrarily complex constraints imposed by multiple instructions (lines 6 and 8 on Listing \ref{lst:extaint}).
Handling those cases with precision thus requires some form of constraint tracking, i.e., symbolic execution or other similar techniques.
This issue is reflected in the literature with works using lightweight symbolic execution to refine dynamic taint propagation \cite{taintpipe, taintse}.}

{In conclusion, taint analysis cannot prove weak control, the lowest level of control in our framework.
It is thus insufficient for evaluating control with any degree of precision.}

\subsection{Verifying Weak and Strong Control with SMT Solvers}

As discussed previously, we need to analyze path constraints in order to properly evaluate control, hence our reliance on symbolic execution.
The following algorithms check Weak and Strong Control for a variable $v$ in a symbolic state $\phi$.

\begin{algorithm}[!htb]
    \caption{$SE_{WC}$}
    \label{alg:sewc}
    \begin{algorithmic}
        \Require $v$ a variable, $l$ a location, $\pi$ a path ending at $l$
        \Ensure returns $true \iff WC_\pi(v,l)$
        \State $\phi \gets SE(\pi)$
        \State $\phi', v' \gets duplicate(\phi, v)$
        \State \Return $sat(\phi \land \phi' \land v \neq v')$
    \end{algorithmic}
\end{algorithm}

Algorithm \ref{alg:sewc} checks $WC$ by checking satisfiability for two different values of $v$.
It is correct and complete for $WC_\pi$.

\begin{proposition}
    $SE_{WC}(v,l,\pi) \iff WC_\pi(v,l)$
\end{proposition}

\begin{algorithm}[!htb]
    \caption{$SE_{SC}$}
    \label{alg:sesc}
    \begin{algorithmic}
        \Require $v$ a variable, $l$ a location, $\pi$ a path ending at $l$
        \Ensure returns $true \iff SC_\pi(v,l)$
        \State $\phi \gets SE(\pi)$
        \Procedure{SC}{$v, l, \phi, E$} \Comment{$E \subseteq Dom(v)$}
            \State $\phi' \gets \exists$ $y \in E$: $\forall$ $x$, $\phi(x) \Rightarrow val(\phi(x),v) \neq y$
            \State \Return $sat(\phi')$, $val(\phi', y)$
        \EndProcedure
        \State $res, y \gets$ \Call{SC}{$v, l, \phi, Dom(v)$} \Comment{$y$: counterex. if not $nil$}
        \State \Return $\lnot res$
    \end{algorithmic}
\end{algorithm}

Algorithm \ref{alg:sesc} checks $SC$ by searching for a counterexample, i.e., an infeasible value.
It is correct and complete for $SC_\pi$.

\begin{proposition}
    $SE_{SC}(v,l,\pi) \iff SC_\pi(v,l)$
\end{proposition}

\begin{algorithm}[!htb]
    \caption{$SE_{SC}|_E$}
    \label{alg:sesce}
    \begin{algorithmic}
        \Require $v$ a variable, $l$ a location, $\pi$ a path ending at $l$, $E \subseteq Dom(v)$
        \Ensure returns $true \iff SC_\pi|_E(v,l)$
        \State $\phi \gets SE(\pi)$
        \State $res, y \gets$ \Call{SC}{$v, l, \phi, E$} \Comment{$y$: counterexample if not $nil$}
        \State \Return $\lnot res$
    \end{algorithmic}
\end{algorithm}

Similarly, Algorithm \ref{alg:sesce} checks $SC$ over $E \subseteq Dom(v)$ ($SC|_E$) by limiting the search for an infeasible value to $E$.

\myparagraph{Solver requirements} 
While $SE_{WC}$ can be verified by any SMT solver, $SE_{SC}$ requires quantifier support.
In addition, obtaining the counterexample $y$ requires the ability to extract models in quantified SMT formulas. 

\subsection{Extracting Domains of Control with Shrink and Split}

\label{sec:algsns}

Giving a simple characterization of $DoC_\pi(v,l)$ as a set of intervals would greatly help further analysis, for example allowing to compute weighted quantitative control as discussed in Section \ref{sec:def}.
To achieve this, we propose the \textit{Shrink and Split} approach detailed in Algorithm \ref{alg:sas}, which consists in repeatedly \textit{shrinking} and \textit{splitting} $Dom(v)$ until we reach $DoC_\pi(v,l)$.

\begin{algorithm}[!htb]
    \caption{$SE_{S\&S}$}
    \label{alg:sas}
    \begin{algorithmic}
        \Require $v$ a variable, $l$ a location, $\pi$ a path ending at $l$
        \Ensure returns $DoC_\pi(v,l)$
        \State $\phi \gets SE(\pi)$
        \Procedure{S\&S}{$v, l, \phi, i, c$} \Comment{$c$: additional constraints}
            \newline\centerline{{- Shrinking -}}
            \State $\phi \gets \phi \land v \in i$
            \State $lo \gets min(val(\phi, v))$
            \State $hi \gets max(val(\phi, v))$
            \State $i \gets [lo;hi]$
            \State $\phi \gets \phi \land v \in i$
            \newline\centerline{{- Checking for Strong Control -}}
            \State $sc, y \gets$ \Call{SC}{$v, l, \phi, \{y \in i / c(y)\}$}
            \If{$sc$}
                \newline\centerline{{- No need to split -}}
                \State \Return $i$
            \Else
                \newline\centerline{{- Splitting around $y$ -}}
                \State \Return \Call{S\&S}{$v, l, \phi, [lo;y[, c$} \\ \hspace{\algorithmicindent} \hspace{\algorithmicindent} \hspace{\algorithmicindent} \hspace{\algorithmicindent} $\cup$ \Call{S\&S}{$v, l, \phi, ]y;hi], c$}
            \EndIf
        \EndProcedure
        \State \Return \Call{S\&S}{$v, l, \phi, Dom(v), True$}
    \end{algorithmic}
\end{algorithm}

\myparagraph{Shrinking}
Intervals are \textit{shrunk} by finding their \textit{feasible bounds}.
To achieve this, we need a \textit{min} and a \textit{max} directive.
In practice we use the SMT solver Z3's optimization modules, which rely on MaxSMT solvers in the case of bitvectors \cite{z3_maxsat}.
Alternatively, feasible bounds can be determined using a binary-search-like method, requiring $\mathcal{O}(log(|Dom(v)|))$ calls to the solver. 
As a result, infeasible values outside of those feasible bounds are eliminated.

\myparagraph{Checking for Strong Control}
We then check whether $v$ is strongly controlled over the shrunk interval $i$.
An affirmative result indicates that $i$ is a subset of $DoC(v,l)$, otherwise we know of at least one infeasible value $y$.

\myparagraph{Splitting}
If infeasible values remain in $i$, we \textit{split} it around $y$ and repeat the $S\&S$ process on both halves.

\myanonparagraph
Since variables are represented as finite bitvectors, all infeasible values are eventually eliminated and we are left with a union of intervals exactly matching $DoC(v,l)$.

\begin{proposition}
    $DoC_\pi(v,l) = SE_{S\&S}(v, l, Dom(v), \pi)$
\end{proposition}

\myparagraph{Practical Limits}
In practice, we observe that proving strong control or finding a counterexample can fail, due to the complexity of constraints, time-outs, solver limitations or bugs...
In this case, weak control at least is guaranteed on the current interval as feasible bounds were found.
{In the worst case, $S\&S$ may only be able to show weak control over the entire domain, although this never happens in our experiments.}

In addition, $S\&S$ has an obvious flaw: when the domains of control contains a lot of holes, the number of splits required may be impractically large.

\myparagraph{Mitigation 1: limited splitting}
One way to avoid excessive splitting is to set a limit and interrupt $S\&S$ when it is reached.
This can be a good solution as intermediate results get more precise over time, although it sacrifices exactness.

\myparagraph{Mitigation 2: fixed bits} Another possible mitigation is to identify fixed bits and take them into account with an additional constraint, preventing splitting caused by them.

\begin{algorithm}[!htb]
    \caption{$SE_{S\&SFB}$: $SE_{S\&S}$ with a fixed bits constraint}
    \label{alg:snsfb}
    \begin{algorithmic}
        \Require $v$ a variable, $l$ a location, $\pi$ a path ending at $l$
        \Ensure returns $DoC_\pi(v,l)$
        \State $\phi \gets SE(\pi)$
        \State $\phi', v' \gets duplicate(\phi, v)$
        \State $\phi'', v'' \gets duplicate(\phi, v)$
        \State $\phi'' \gets \phi \land \phi' \land \phi'' \land mask = \sim (v'$ $\hat{}$ $v'')) $ \\ \hspace{\algorithmicindent} $\land bits = v' \& v''$ \Comment{(a)}
        \State $\phi'' \gets \phi'' \land (\lnot \exists x$: $\phi''(x) $\\ \hspace{\algorithmicindent} $\land val(\phi''(x), v)$ $\hat{}$ $mask \neq bits))$ \Comment{(b)}
        \If{$sat(\phi'')$}
            \State $mask \gets val(\phi'', mask)$
            \State $bits \gets val(\phi'', bits)$ 
        \Else
            \State $mask \gets 0$
            \State $bits \gets 0$
        \EndIf
        \State \Return \Call{S\&S}{$v, l, \phi, Dom(v), y \mapsto y$ $\hat{}$ $mask = bits$}
    \end{algorithmic}
\end{algorithm}

Algorithm \ref{alg:snsfb} shows how we compute fixed bits and incorporate them into $S\&S$.
$(a)$ and $(b)$ ensure that $mask$ corresponds to at least and at most all fixed bits respectively, while $bits$ contains their value.

There exist corner cases where two feasible values always have more common bits than the number of overall fixed bits, in which case this algorithm fails with $mask$ and $bits$ null.
However we expect such cases to be rare.
Alternatively we could evaluate each bit individually, however we observe that our approach is more efficient in practice.

Once the fixed bits constraint is determined, we incorporate it within $S\&S$ by checking strong control over values satisfying it in the shrunk interval.
This implicitly eliminates holes and prevents excessive splitting.

This variant of $S\&S$ is also evaluated in Section \ref{sec:xps} and shown to improve precision with no noticeable overhead.

\myparagraph{Other possible mitigations}
Other types of constraints such as congruences or polyhedra could be similarly considered in lieu of fixed bits. 
This is left as future investigation.  

\myparagraph{Approximation}
When an exact results cannot be computed, Shrink and Split still gives an over- and under-approximation of the domains of control, respectively by taking all resulting intervals and only those where strong control is proven.

\myparagraph{Solver requirements} 
Both $SE_{S\&S}$ and $SE_{S\&SFB}$ incur the same solver requirements as $SE_{SC}$, with the addition of the \textit{min} and \textit{max} directives.

\myparagraph{Comparison with Newsome et al.'s algorithm \cite{quantinfl}}
Our $S\&S$ algorithm bears some similarities with Newsome et al.'s feasible value set estimation algorithm.
Both give a representation of what we call domains of control as a set of intervals, with a distinct splitting operation.
Both also require $min$ and $max$ solver directives.
However their algorithm splits intervals around random feasible values, which is fairly inefficient.
Finally, they perform statistical sampling and compute a confidence interval on the density of each interval rather than proving strong control, leading to weaker guarantees.

On the other hand, our algorithm requires solving quantified SMT formulas and optimization queries, while theirs only issues standard quantifier-free SMT queries.

\subsection{Conclusion}

\begin{table}[!htb]
    \caption{Comparison of analysis methods for attacker control}
    \label{tab:methcomp}
    \resizebox{\columnwidth}{!}
    {
        \begin{tabular}{lcccccc}
            \hline
            Algorithm & TA & $SE_{WC}$ & $SE_{SC}$ & $SE_{PMC}$ & Newsome & $SE_{S\&S}$ \\
            \hline
            underlying  & - & QF SMT & Q SMT & PMC & QF SMT, & Q SMT, \\
            problem(s)  &   &        &       &     & min / max & min / max \\
            \hline
            $\lnot WC$ & \tick    & \tick     & \cross     & \tick  & \tick    & \tick     \\
            $WC$ & \cross    & \tick     & \tick *    & \tick & \tick    & \tick     \\
            $SC$ & \cross    & \cross    & \tick     & \tick & \meh    & \tick     \\
            $QC$ & \cross    & \cross    & \cross    & \tick & \meh     & \tick     \\
            $DoC$ & \cross    & \cross    & \cross    & \cross & \meh    & \tick     \\
            $wQC$ & \cross & \cross & \cross & \cross & \meh & \tick \\
            \hline
        \end{tabular}
    }
    \par\smallskip
    \scriptsize QF SMT $=$ Quantifier Free SMT, Q SMT $=$ Quantified SMT, PMC $=$ Projected Model Counting, min / max $=$ directives for min and max values.
    \tick{} means that an exact result can be given, \meh{} that only approximation can be given and \cross{} that no result can be given.\\
    * implicit when strong control is proven, but limited
\end{table}

As summed up in Table \ref{tab:methcomp}, we have shown that taint analysis cannot give much guarantees on attacker control.
In comparison, all of our control properties can be proven or measured using algorithms based on symbolic execution.
In particular, $SE_{S\&S}$ is able to compute domains of control, from which all other notions are derived.

\section{Experimental Validation}

\label{sec:xps}

We will now evaluate our approach and compare the characteristics of Shrink and Split with the other algorithms.

\subsection{Implementation}

We have implemented all the different techniques mentioned so far in a new generic binary-level dynamic analysis engine designed for flexibility and quick prototyping. 
We build on the Intel PIN binary instrumentation framework \cite{pin}, the BINSEC symbolic execution engine \cite{binsec, binsec_se} and several external solvers.  
Our prototype is currently limited to  x86\_64 due to the use of PIN tools.  
Additionally: 

\begin{itemize} 

\item We perform taint analysis at the BINSEC IR level, allowing us to precisely track intra-instruction data-flows; 

\item SMT solving is performed via a portfolio approach with Z3 \cite{z3}, Bitwuzla \cite{bitwuzla} in conjunction with various SAT solvers and Q3B \cite{q3b}; 

\item We support the exact PMC solvers d4 \cite{d4} and ganak \cite{ganak} and the approximate solver approxmc \cite{approxmc_CMV16, approxmc_SM19, approxmc_cav20} with $\mathcal{P}(result = truth \pm 20\%) = 0.95$ as a guarantee; 

\item We set a limit of k=100 splits for S\&S\footnote{The study in Appendix \hyperref[tab:appsumsplitres]{V} shows a negligible impact of k on performance and precision above a certain threshold.}; 

\item We reimplemented Newsome et al's algorithm based on the details provided in their paper \cite{quantinfl}.

\end{itemize}

The choice of precise yet architecture-specific binary-level analysis limits approximation in path constraints and thus ensures that the precision of control algorithms can be properly evaluated.
However it is not an integral part of our method as any other way of deriving path constraints can be used, such as source level symbolic execution.

\myparagraph{Automatically detecting and analyzing out-of-bounds memory accesses}
We implemented an analysis based on taint which tracks pointers toward stack, heap and global objects.
The taint information contains the bounds of the object and allows to check for violations during memory accesses.
This allows to automatically identify and analyze the target variables for such vulnerabilities.

\subsection{Benchmark}

\label{sec:bench}

Our benchmark is split into a set of programs with well-understood vulnerabilities and another, more realistic one.

\begin{table}[!htb]
    \caption{Real-world vulnerabilities from the ground-truth benchmark B1 (see Appendices \ref{tab:app_gt_bench} and \ref{fig:app_gt_doms} for more details)}
    \label{tab:knownrealbench}
    \resizebox{\columnwidth}{!}{
    \begin{tabular}{l l c r r}
        \hline
        Program & \multicolumn{2}{l}{Vulnerability} & \multicolumn{2}{l}{Executed Instructions}\\
                & name          & type & Symbolic & Total \\
        \hline
        libjpeg & cve-2023-37837 & OOBR & 56 & 261k \\
        libsndfile & cve-2021-3246 & OOBW & 432 & 47k \\
        mongoose & cve-2019-19307 & IOF & 104 & 67k \\
        u-boot & cve-2019-14192 & OOBW & 38 & 4k \\
        u-boot & cve-2019-14202 & OOBW & 38 & 4k \\
        u-boot & cve-2022-30790 & OOBW & 161 & 3k \\
        u-boot & cve-2022-30790-2* & OOBW & 34 & 3k \\
        u-boot & cve-2022-30552 & OOBW & 91 & 2k \\
        openssl & heartbleed & OOBR & 33 & 165 mil. \\
        faad2 & cve-2021-26567 & CFH & 14 & 12k \\
        perl-dbi & cve-2020-14393 & CFH & 2491 & 60 mil. \\
        sdop & cve-2024-41881 & CFH & 60 & 657k \\
        xfpt & cve-2024-43700 & CFH & 271 & 179k \\
        mjs & cve-2023-43338 & CFH & 99 & 51k \\
        \hline
    \end{tabular}
    }
    {
    \par\smallskip
    \scriptsize
    OOBR/W $=$ Out-Of-Bounds Read / Write, IOF $=$ Integer OverFlow, CFH $=$ Control-Flow Hijacking
    \par
    *side effect of cve-2022-30790
    \par
    }
\end{table}

\myparagraph{Ground-truth benchmark (B1)}
This benchmark is composed of the $14$ real-world vulnerabilities from Table \ref{tab:knownrealbench}, plus $8$ synthetic examples from Newsome et al. \cite{quantinfl} and  $17$ new ones. 
The bugs in these programs were manually analyzed and are thus well understood.
\textit{The purpose of this benchmark will be to demonstrate the correctness of our approach.}

The real-world vulnerabilities in this benchmark were selected from the literature (\textit{cve-2019-19307} and \textit{cve-2019-14192} \cite{robreach}, heartbleed) and CVE databases.
Regarding the latter, our requirements for open-source and reproducibility are {surprisingly rarely fulfilled}.
The size of this benchmark is further limited by the need to establish ground truths manually and the lack of reuseable examples from the literature.

\begin{table}[!htb]
    \caption{Realistic benchmark B2 (see also Appendix \ref{tab:app_real_bench})}
    \label{tab:magmashort}
    \resizebox{\columnwidth}{!}{
    \begin{tabular}{l l l l}
        \hline
        Program & \# of Vulnerabilities & \multicolumn{2}{l}{Executed Instructions} \\
                &                       & Symbolic & Total (million) \\
        \hline
        poppler & 5 & 0 - 6671 & 5 - 15 \\
        openssl & 9$^1$ & 91 - 5019 & 1 - 30 \\
        libtiff & 5$^2$ & 274 - 1086 & 0.1 - 1 \\
        libxml & 4 & 0 - 417 & 0.1 - 7 \\
        php & 1 & 196 & 6 \\
        libpng & 1 & 0 & 0.02 \\
        sqlite3 & 1 & 0 & 0.25 \\
        \hline
    \end{tabular}
    }
    {
        \par\smallskip
        \scriptsize
        $^1$including 6 variants not caught by Magma's ground truth oracles\\
        $^2$including 2 variants with different capabilities
        \par
    }
\end{table}

\myparagraph{Realistic benchmark (B2)}
This benchmark is composed of 26 out-of-bounds memory vulnerabilities from the Magma state-of-the-art fuzzing benchmark \cite{magma}, in programs such as openssl, libtiff and libxml (see Table \ref{tab:magmashort}).
They include buffer overflows, use-after-frees and various other invalid memory accesses.
We picked these vulnerabilities based on whether they were triggered during the original evaluation of Magma.

These vulnerabilities were analyzed based solely on the available reproducing input and ASAN reports. 
Contrary to B1, no manual instrumentation was performed, hence the increase in realism toward in-the-field analysis of large, complex programs.
Yet we lack a clear ground-truth for 
these vulnerabilities.  
\textit{The purpose of this benchmark will thus be to demonstrate the practicality of our approach.}

\subsection{Research Questions}

We structure the experimental evaluation of our bug prioritization approach around the following research questions:

\myparagraph{RQ1}
How precise is Shrink and Split and how does it compare to other algorithms?

\myparagraph{RQ2}
How scalable is Shrink and Split and how does it compare to other algorithms?

\myparagraph{RQ3}
How effective is our approach at prioritizing bugs compared to others?

\myparagraph{RQ4}
How does our approach fare in realistic end-to-end scenarios?

\myanonparagraph We chose a limit of 100 splits for S\&S in our experiments. 
See Appendix \ref{tab:appsumsplitres} for a study of the split limit's impact.

\subsection{RQ1: Precision}

\label{sec:rq1}

We first compare the precision of S\&S against other qualitative and quantitative methods.

\begin{figure}[!htb]
    \includegraphics[width=\columnwidth]{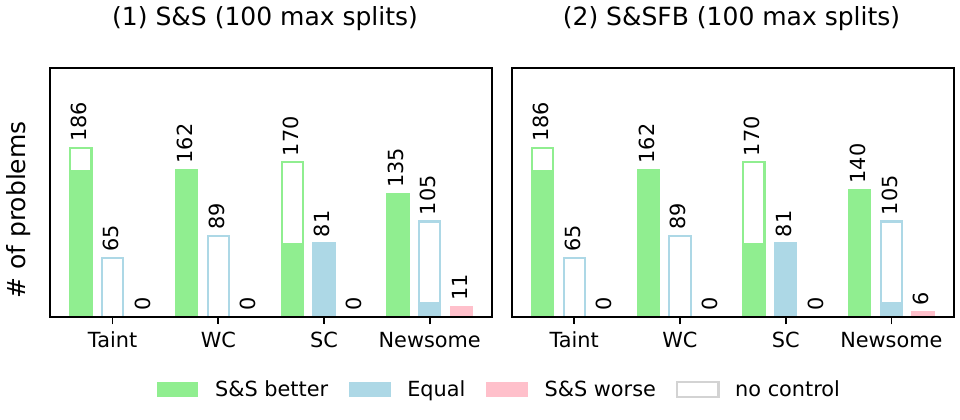}
    \par\smallskip\small\textit{Improvements in cases without control are due to 24 false positives for taint analysis and the fact that non-SC does not mean absence of control for SC.}

    \caption{Precision of domains of control with S\&S compared to other applicable algorithms}
    \label{fig:preccompdoms}
\end{figure}

\begin{figure}[!htb]
    \includegraphics[width = \columnwidth]{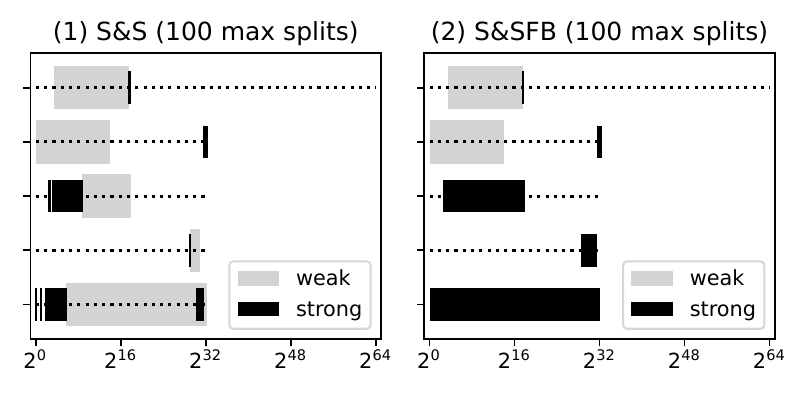}
    {
        \par\vspace{-.5cm}\small\textit{
        From top to bottom: cve-2021-3246.of\_size, cve-2022-30790.wlen, cve-2022-30790\_2.woff2, grub.canary, mul.j.
        See Appendix \ref{fig:app_gt_doms} for more examples.
        {The dotted lines represent maximal value ranges based on variable sizes.}
        }
        \par
    }
    \caption{Examples of approximate results with S\&S}
    \label{fig:snsappdoms}
\end{figure}

\paragraph{Qualitative methods.} The results are compared based on the inclusion of the returned domains of control, with smaller ones being considered better. 
Figure \ref{fig:preccompdoms} shows that S\&S is more precise than other applicable algorithm for measuring domains of control in most instances.
In particular, it is strictly better than Newsome et al's algorithm on 135 (140 with S\&SFB) cases out of 251, and worse in only 11 cases  (6 with S\&SFB).  
This is partly due to the loss of precision when the domains contain too many holes, which is mitigated by S\&SFB (see Figure \ref{fig:snsappdoms}). 
On the other hand, S\&S always improves upon WC when there is control, i.e., in 162 cases out of 251.
It also beats taint analysis in these instances -- note that tainting suffers from  24 false positives here. 
Finally, S\&S can show absence of control while SC cannot, and also improves upon the latter in 81 cases with control. 

\begin{figure}[!htb]
    \includegraphics[width=\columnwidth]{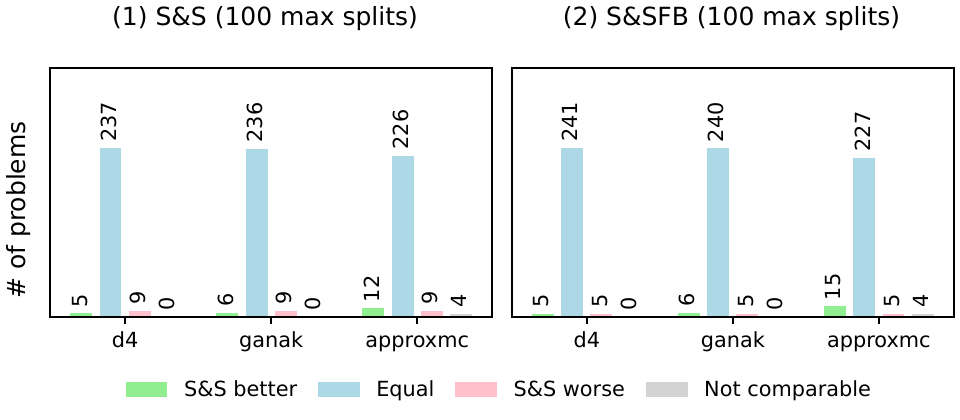}
    \par\smallskip\small\textit{Results from approxmc cannot be compared without an exact count.}\par
    \caption{Precision of quantitative control with S\&S compared to model counters}
    \label{fig:preccompqc}
\end{figure}

\paragraph{Quantitative methods.} We compare S\&S to counting methods by reducing the domains to a single value count. 
It must be clear that S\&S provides more information than quantitative methods by design, yet this comparison is restricted to counting only.
Figure \ref{fig:preccompqc} details how S\&S fares against d4 and ganak, as well as the approximate PMC solver approxmc.
Overall S\&S performs well in most instances, while S\&SFB improves precision in 4 cases.
Both algorithms also beat d4 and ganak on some problems due to them timing out.
Finally, S\&S and S\&SFB beat approxmc in 12 and 15 cases.

\medskip

\concl{\textbf{Conclusion RQ1}. While $S\&S$ may lose precision in some cases, it overall improves upon Newsome et al.'s algorithm and is even competitive against PMC solvers.}

\subsection{RQ2: Performance}

\label{sec:rq2}

Bug-prioritization methods can afford to be slower than bug-detection methods, as they only need to analyze buggy executions.
Nevertheless, scalability always benefits usability.

\begin{table}[!htb]
    \caption{Cumulative runtimes for each algorithm on B1$+$B2}
    \label{tab:runtimes}
    
\begin{tabularx}{\columnwidth}{lCCr}
\hline
 Algorithm               & Notion &    Time-outs$^1$      &   Runtime$^2$ \\
\hline
 Taint                   & WC     &         0 & $<$ 0.1s \\
 WC             & WC     &         0  & 3s \\
 SC           & SC     &        0 &  46s\\
\hline
 d4                      & QC     &         5 &    36m \\
 ganak                   & QC     &         6 &    33m\\
 approxmc                & QC     &          1 &    15m\\
\hline
 Newsome                 & DoC    &          1 &   2h09m \\
 S\&S$^3$                    & DoC    &       0 &   11m\\
 S\&SFB$^3$                  & DoC    &        0 &  11m\\
\hline
\end{tabularx}
    \par\smallskip\scriptsize$^1$5 minutes --- $^2$including time-outs --- $^3$100 maximum splits
    
\end{table}

Table \ref{tab:runtimes} shows that S\&S is  competitive in terms of performance, with no time-outs and a total runtime of $11$ minutes for both variants, beating its direct competitor Newsome et al.'s algorithm by a very large margin (2h09, 1 TO), with an average speedup of $73\times$ for S\&S and $77\times$ for S\&SFB\footnote{without the $5\%$ top and bottom outliers}.
This also shows that the additional fixed bits constraints does not meaningfully degrade performance.
Counting approaches also suffer from time-outs.

\medskip 

\concl{\textbf{Conclusion RQ2}. Both $S\&S$ and $S\&SFB$ are significantly faster than the existing state-of-the-art and more robust than projected model counting algorithms.}

\subsection{RQ3: Bug priorization}

\label{sec:rq3}

So far we have evaluated the precision and scalability of Shrink and Split in relation to expectations and other algorithms.
We will now show how our approach allows to precisely prioritize vulnerabilities via weighted quantitative control, with a case study on the $8$ real-world buffer out-of-bounds vulnerabilities and our motivating example from our ground-truth benchmark B1.
{In addition, we also present a case study on $5$ real-world control-flow hijack primitives (code pointer corruption).
Its purpose is to illustrate the more simple data control problem.}

\begin{table}[!htb]
    \caption{Using control to rate vulnerabilities}
    \label{tab:scores}
    \par\small\textit{
    \textbf{Out-Of-Bounds (memory range problem).}
    We rate control over memory ranges as the sum of the QC or wQC scores for offsets and sizes, multiplied by variable sizes.
    Non-out-of-bounds values are excluded and offsets are relative to the closest relevant bound for OOB write vulnerabilities.
    For wQC, we use the weight function $\omega : x \mapsto \frac{1}{ln(2)x}$ as in Section \ref{sec:wqc}.
    Other similar functions yield equivalent results, see Appendix \ref{tab:appwfunscores} for more details.\\
    {\textbf{Control-Flow Hijacking (data problem).}
    We rate control over a corrupted code pointer.
    Since variable sizes are always 8 bytes, scores are normalized.
    When string parsing is involved, analysis takes into account individual byte constraints.
    For wQC, we give a null weight to invalid addresses, i.e., those corresponding to non-executable or unmapped memory.
    {In particular, in 64-bit Linux, addresses with two non-null high bytes are invalid in user-space.}
    }
    }
    \par\medskip
    Score categorization:\\
    \Smiley[1][green] (low capabilities): CVSS $<5$, OOB $< 1$, CFH $< 0.01$\\
    \Neutrey[1][yellow] (medium capabilities): CVSS $<7$, OOB $<10$, CFH $<0.1$\\
    \Sadey[1][red!50!white] (high capabilities): CVSS $\geq7$, OOB $\geq10$, CFH $\geq0.1$
    \medskip\par
    \resizebox{\columnwidth}{!}{
    \begin{tabular}{l H H l c l c l c |c}
        \hline
        Vulnerability & Offset & Size & \multicolumn{2}{l}{CVSS} & \multicolumn{2}{l}{QC Score} & \multicolumn{2}{l |}{wQC Score} & Human$^2$ \\
        \hline
        \multicolumn{10}{c}{OOB writes} \\
        \hline
        motex1 & \textit{fixed} & of\_wsize & - & & 5.32 & \Neutrey[1.5][yellow] & 0 & \Smiley[1.5][green] & \Smiley[1.5][green] \\
        motex2 & \textit{fixed} & of\_wsize & - & & 5.32 & \Neutrey[1.5][yellow] & 5.36 & \Neutrey[1.5][yellow] & \Neutrey[1.5][yellow] \\
        
        cve-2021-3246 & \textit{fixed} & of\_size & 8.8 & \Sadey[1.5][red!50!white] & 17.6 & \Sadey[1.5][red!50!white] & 14.14 & \Sadey[1.5][red!50!white] & \Sadey[1.5][red!50!white] \\
        cve-2019-14192 & \textit{fixed} & of\_wsize & 9.8 & \Sadey[1.5][red!50!white] & 15.97 & \Sadey[1.5][red!50!white] & 15.97 & \Sadey[1.5][red!50!white] & \Sadey[1.5][red!50!white] \\
        cve-2019-14202 & \textit{fixed} & of\_wsize & 9.8 & \Sadey[1.5][red!50!white] & 15.97 & \Sadey[1.5][red!50!white] & 15.97 & \Sadey[1.5][red!50!white] & \Sadey[1.5][red!50!white] \\
        cve-2022-30790 & of\_woff(\_2) & of\_wsize(\_2) & 7.8 & \Sadey[1.5][red!50!white] & 15.6 & \Sadey[1.5][red!50!white] & 0.51 & \Smiley[1.5][green] & \Smiley[1.5][green] \\
        cve-2022-30552 & of\_woff & of\_wsize & 5.5 & \Neutrey[1.5][yellow] & 15.6 & \Sadey[1.5][red!50!white] & 0.51 & \Smiley[1.5][green] & \Smiley[1.5][green] \\
        cve-2022-30790-2 & of\_woff2 & \textit{fixed} & - & & 15.94 & \Sadey[1.5][red!50!white] & 2.37 & \Neutrey[1.5][yellow] & \Neutrey[1.5][yellow] \\
        \hline
        \multicolumn{10}{c}{OOB reads}\\
        \hline
        cve-2023-37837 & read\_off & \textit{fixed} & 6.5 & \Neutrey[1.5][yellow] & 16 & \Sadey[1.5][red!50!white] & 16 & \Sadey[1.5][red!50!white] & \Sadey[1.5][red!50!white] \\
        heartbleed & \textit{fixed} & payload\_size & 7.5 & \Sadey[1.5][red!50!white] & 16 & \Sadey[1.5][red!50!white] & 16 & \Sadey[1.5][red!50!white] & \Sadey[1.5][red!50!white] \\
        \hline
        \multicolumn{10}{c}{CFH}\\
        \hline
        cve-2021-26567 & & & 7.8 & \Sadey[1.5][red!50!white] & $0.97$ & \Sadey[1.5][red!50!white] & 0 & \Smiley[1.5][green] & \Smiley[1.5][green]\\
        cve-2020-14393 & & & 7.1 & \Sadey[1.5][red!50!white] & $0.99$ & \Sadey[1.5][red!50!white] & 1 & \Sadey[1.5][red!50!white] & \Sadey[1.5][red!50!white]\\
        cve-2024-41881 & & & 8.8 & \Sadey[1.5][red!50!white] & $0.94$ & \Sadey[1.5][red!50!white] & 0 & \Smiley[1.5][green] & \Smiley[1.5][green] \\
        cve-2024-43700 & & & 7.8 & \Sadey[1.5][red!50!white] & $0.78$ & \Sadey[1.5][red!50!white] & 0 & \Smiley[1.5][green] & \Smiley[1.5][green]\\
        cve-2023-43338 & & & 9.8 & \Sadey[1.5][red!50!white] & $10^{-5}$ & \Smiley[1.5][green] & 1 & \Sadey[1.5][red!50!white] & \Sadey[1.5][red!50!white]\\
        \hline
        Total Correct & & & & $6/12$ & & $7/15$ & & $15/15$ \\
        \hline
    \end{tabular}
    }
    {
        \par\smallskip
        \scriptsize
        $^1$$7$ for CVSS --- {$^2$manual analysis performed by us}
        \par
    }
\end{table} 

\begin{figure}[!htb]
    \centering
    \includegraphics[width=\columnwidth]{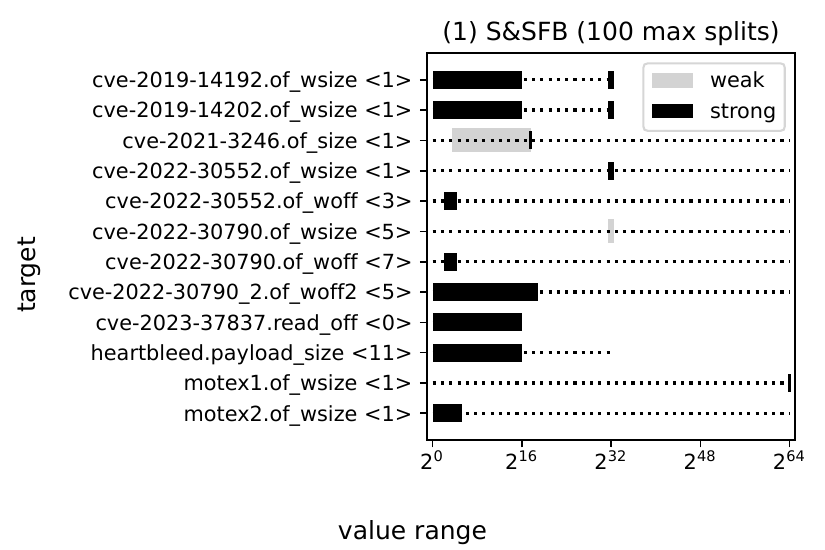}
    \caption{Domains of control for the OOB vulnerabilities}
    \label{fig:rq3doms}
\end{figure}

\myparagraph{Distinguishing between different vulnerabilities}
Table \ref{tab:scores} details the CVSS, QC and wQC scores for each of the vulnerabilities in our study.
These results show that weighted quantitative control, and by extension domain of control analysis, is able to clearly differentiate between vulnerabilities with different capabilities, such as \textit{motex1} and \textit{motex2} (resp. vulnerabilities a and b from Section \ref{sec:motex}) or \textit{cve-2022-30552} and \textit{cve-2019-14192}, with the former only allowing for very large write sizes (see Figure \ref{fig:rq3doms}) and thus being less dangerous.

\myparagraph{Identical capabilities}
Additionally, domains of control allow to recognize vulnerabilities sharing the same capabilities, i.e., those with the same target variables and domains of control, such as \textit{cve-2019-14192} and \textit{cve-2019-14202} or \textit{cve-2022-30790} and \textit{cve-2022-30552} (see Figure \ref{fig:rq3doms}).

\myparagraph{Correcting human analysis}
{In the case of \textit{cve-2022-30790}, we found previous human analysis to be incorrect. 
It was discovered together with \textit{cve-2022-30552} \cite{nccgroup} and was thought to grant an arbitrary write primitive by overwriting metadata in a linked list. 
In contrast, the other can only be used for DOS attacks due to a very large write size caused by an integer underflow.
However, our results contradict this interpretation as both vulnerabilities are found to have the same capabilities, matching the analysis of \textit{cve-2022-30552}.
Inspecting the code reveals that some security checks prevent most out-of-bounds writes, which must have been missed by previous evaluators (see Appendix \hyperref[app:ubootdefrag]{VIII}).
Using our tool could have helped them realize their mistake and thus improve human analysis.}

\myparagraph{Comparing control-flow hijacking primitives}
{The main point of interest regarding these vulnerabilities is the fact that QC is particularly misleading.
This is due to the fact that the two upper bytes of the corrupted code pointer must be set to zero in order to prevent a crash.
As a result, vulnerabilities with high QC such as \textit{cve-2021-26567} and \textit{cve-2024-41881} are not exploitable since the data overwriting the code pointer comes from a string, hence bytes cannot be null.

In contrast, \textit{cve-2023-43338} has low QC but is very controllable since any value between $0$ and $2^{48} - 1$ can be obtained (see Appendix \hyperref[fig:app_gt_doms]{III}).}

\myparagraph{Comparison with other methods}
CVSS and quantitative control scores are not so useful as differences are too minor to be meaningful.
This is to be expected for the former as it is only a loose indicator assuming worst case.
On the other hand, quantitative control {cannot account for differing value threat levels}.
Weak and strong control alone are also not very useful as in most cases control is weak but not strong.
{Overall, weighted QC scores are the only ones to match our expectations, with 12/12 matches for the vulnerabilities with CVSS scores, against 6/12 for the latter and QC.}

\medskip

\concl{\textbf{Conclusion RQ3.} Our approach allows to prioritize vulnerabilities more precisely than the existing state-of-the-art in our case study. It would also have improved human analysis for \textit{cve-2022-30790}.}

\subsection{RQ4: Realistic end-to-end scenarios}

\label{sec:rq4}

\begin{figure}[!htb]
    \includegraphics[width=\columnwidth]{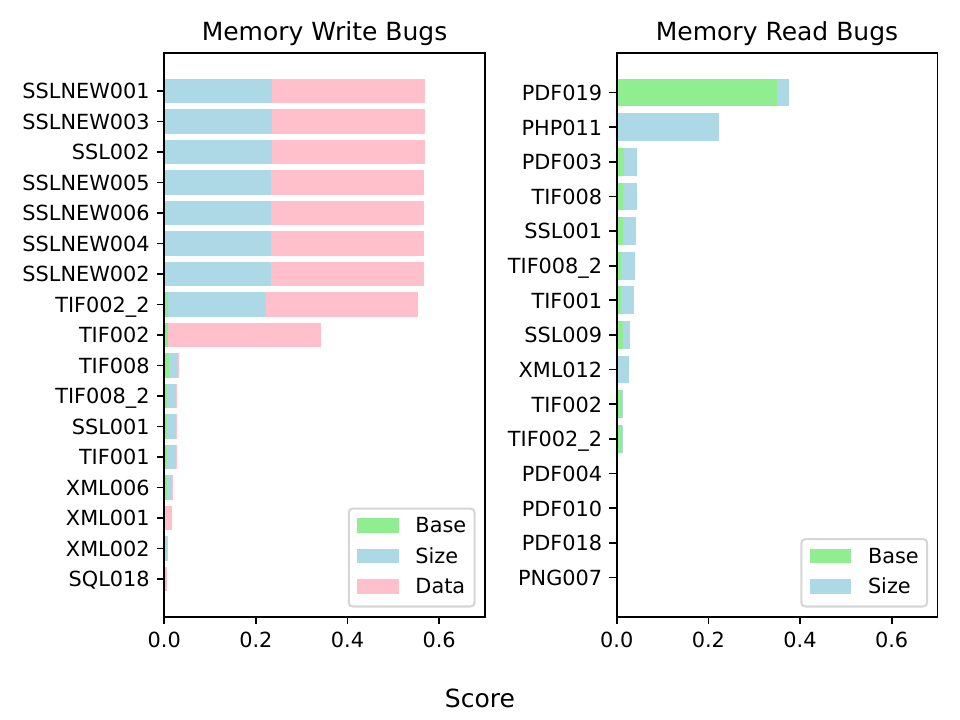}
    \small\textit{
    Scores are computed for {base addresses} and sizes using weighted quantitative control {biased toward local OOBs}. 
    Our weight function is $x \mapsto \frac{1}{ln(2)d(x)}$ with $d$ the distance between $x$ and the nearest buffer bound {for addresses} or minimum out-of-bounds size {for sizes}.
    Data scores are given as an average of quantitative control over the first eight written bytes.
    }
    \caption{OOB capability scores for the Magma bugs}
    \label{fig:magma_scores}
\end{figure}

To demonstrate the practicality of our approach, we automatically analyzed and rated the Magma vulnerabilities from benchmark B2, starting from a single input only and with practically no human effort. 
Figure \ref{fig:magma_scores} displays the resulting scores, which clearly differentiate vulnerability capabilities. 
Overall, our wQC scores match expectations and allows us to derive interesting knowledge of the vulnerabilities (see Appendix \hyperref[app:rq4interp]{IX} for a more in-depth discussion). 

\myparagraph{OOB writes} We observe a correlation with the 11 available CVSS scores, with 9 out of 11 close matches with our wQC score and a clear distinction between highly and less controllable vulnerabilities. 
Regarding the two mismatches, \textit{SSL001} and \textit{TIF001}, we show that despite their high CVSS scores,  their capabilities are actually very limited, since they consists in repeated non-contiguous single byte writes, with no control over any parameters. 

We also observe a similarity between the \textit{SSLNEW} vulnerabilities' capabilities and those of \textit{SSL002}.
The fact that they are fixed by the same patch suggests that they are indeed most likely derived from the same bug.

\myparagraph{OOB reads} We observe the same trend as OOB writes, although some CVSS scores are influenced by the occurrence of OOB writes in the same execution, complicating the interpretation. 
Additionally, our analysis suggests that \textit{PDF010}, which does not have an associated CVE, has the same capabilities as three other vulnerabilities with CVEs.

\medskip

\concl{\textbf{Conclusion RQ4.} Our approach can be fully automated and still clearly differentiate highly controllable vulnerabilities from others in realistic fuzzing targets.}

\section{Discussion}

\label{sec:disc}

\myfirstparagraph{Handling Other Types of Vulnerabilities} 
\label{sec:disctypes} 
In our experimental validation, we focused on memory out-of-bounds vulnerabilities.
However our method can be applied to other types of vulnerabilities as well. 
In particular, vulnerabilities resulting in control over well-defined data value (data control problem model) can be handled by computing weighted quantitative control over said value with an appropriate weight function.
For example, regarding pointer corruption and assuming a 64-bit architecture, the weight function should attribute a weight of zero to any value greater or equal to $2^{48}$, since those addresses are invalid.

{On the other hand, our tool could also be used to measure information leakages, since they are typically evaluated using data-flow analysis.
For example, in our heartbleed experiment, we analyzed 8 bytes of unencrypted data leaked from a previous server interaction.
Yet indirect leakages (e.g., side-channels) would be difficult to handle.}

Finally, our method may not be applicable to some kinds of vulnerabilities (e.g., DoS caused by infinite loops) and may be difficult to leverage for others (e.g., heap layout manipulation, although we discussed some possibilities in Section \ref{sec:scope}).

\myparagraph{Multi-path analysis and implementation}
While single path analysis simplifies symbolic execution and greatly improves scalability by avoiding path explosion, it can result in a loss of completeness.  
A possible mitigation is to explore additional paths and merge those reaching the same vulnerability, yet it should be done with care in order to avoid any significant performance cost. 
Another solution would be to explore multiple paths separately, then merge the domains of control. 

{Note that our framework, algorithms and general approach does not rely on any particular constraint computation technique, thus other implementations may choose to sacrifice performance for more completeness with multi-path analysis. 
We also investigated how standard symbolic execution optimizations impact our approach.
As expected, constraint relaxation \cite{relax} (over-approximation) yields wider domains (see Appendix \hyperref[tab:app_relax]{XI}) while partial input concretization \cite{dart, concr} (under-approximation) induces tighter ones (see Appendix \hyperref[tab:app_concr]{XII}). 
Both may reduce computation time but also impact bug evaluation. 
A systematic evaluation is left as future work.      
}

\myparagraph{Interpreting results}
While weighted quantitative control can help to automatize vulnerability evaluation, expert insight is still required in order to define a suitable weight function.
This can also arguably be an advantage, as it allows to tailor the method toward the specific needs and constraints of different projects, hardware or environments. 

{On a different note, weak intervals in domains computed by Shrink and Split can be difficult to interpret, as they may have very different densities (see Appendix \hyperref[tab:snscount]{X}). 
Fortunately, our threat classification from Table \ref{tab:scores} is unaffected, but this may not always be the case.}

\myparagraph{Limits of automation}
Our tool  automation is enough to handle many situations, as illustrated in Section \ref{sec:rq4}.
However, it cannot track some implicit properties, such as constraints on the length of strings.
Repeated memory accesses inside loops can also be tricky to properly analyze, as some parameters may overall be tied to the looping condition.
These are universal problems in binary-level analysis. 

\myparagraph{Limits of control for exploitability evaluation}
{Exploitability is a fundamentally hard to capture concept.  
As such, one can only aspire to identify and evaluate some of its aspects, such as control.
This also means that any formal exploitability assessment technique is limited by the scope of its target property.
In our case, control measurements may be too narrow or approximate due to the technical limitations of our implementation, but factors other than control may also be at play, such as a lack of robustness \cite{quantrob}.
}

\myparagraph{Binary vs. source code analysis for prioritization}
{Given the variety of existing platforms, environments and building tools, source-level analysis may appear a good choice for generic bug evaluation. 
Yet, bugs of the types we consider here are often found at the binary level (e.g., ASAN detections during fuzzing) and their  exploitation often depends on low-level concepts such as memory layouts. They are thus difficult to precisely characterize at source level.
}

\section{Related Works}

\label{sec:rw}

\myfirstparagraph{Bug Impact Evaluation}
Manual bug prioritization efforts such as CVSS scores tend to always assume worst case scenarios, regardless of the actual capabilities of vulnerabilities.
In addition, human error is an ever-present risk as illustrated by the case of  \textit{cve-2022-30790} (Section \ref{sec:rq3}).

On the other hand, automated bug prioritization practices usually consist in attributing different priority levels to different types of vulnerabilities. 
In the case of fuzzing, sanitizers such as ASAN \cite{asan} or KASAN \cite{kasan} provide information on bug impacts.
While it works well for coarse-grained prioritization, it does not allow to distinguish between vulnerabilities of the same type. 

Syzscope \cite{syzscope} refines this approach by allowing the execution to continue after a first issue is detected, potentially leading to more serious ones.
While this allows to identify dangerous bugs which would otherwise be considered benign, it still does not allow to distinguish between vulnerabilities of the same type. 

{Evocatio \cite{evocatio} uses targeted fuzzing to discover additional bug capabilities, such as different sizes or offsets of out-of-bounds writes, but does not score nor rank them.} 

{KOOBE \cite{koobe} characterizes out-of-bound reads and writes using symbolic execution in order to identify the most promising ones to automatically build exploits. 
It ranks capabilities only when their associated constraints are either identical, or one of them is a constant value and solution of the other (partial order).  
While useful in an automatic exploit generation setting, this approach is too simple for full-fledged bug prioritization: no two vulnerabilities from Table \ref{tab:scores} could be compared with it. 
Despite its limitations, KOOBE is to our knowledge the closest existing work to offering a generic fine-grained vulnerability capability metric.
}

\myparagraph{ML-based methods}
Existing works attempt to predict the exploitability of vulnerabilities using deep learning techniques \cite{prio_survey, suciu, jiang, bert}.
While these approaches scale well once trained, the lack of transparency in the results as well as the over-reliance on human analysis (bug reports, CVSS scores, etc.) currently hinder their usability.

\myparagraph{Automatic Exploit Generation (AEG)} 
One way to prove the exploitability of a bug is to build an exploit around it.
However, building exploits manually is difficult, hence the interest in  automatic exploit generators (AEG). 
While first attempts focused on   shellcode injection exploits \cite{Heelan,aeg, mayhem}, more recent works focus  on  heap exploitation \cite{Repel, revery, gollum, heaphopper, archeap, shrike, maze}, kernel exploitation \cite{lu, fuze, KEPLER, SLAKE, koobe} or gadget chain synthesis \cite{Q, DOE, DOP, bopc}. 

While a priori close in goals, AEG is not well suited for bug priorization. Indeed, AEG tools tend to be highly specialized, so that   
while building an exploit is the strongest possible proof of exploitability, failure is less conclusive, as the vulnerability could require a  type of exploit not covered by the AEG engine at hand. 
AEG may also involve fuzzing for similar-yet-different vulnerabilities \cite{fuze, koobe, revery}. 

\myparagraph{Robust Reachability} 
Another aspect of the exploitability of bugs besides their security impact is how reliably they can be triggered. 
Girol et al. \cite{robreach,robreachjournal,quantrob} proposed robust reachability to formally capture this notion that  
complements but is orthogonal to bug impact analysis for prioritization. 

\myparagraph{Quantitative Information Flow}
Measuring channel capacity and thus quantitative information flow has mainly been developed for the purpose of quantifying leakage of secret information \cite{heusser, mccamant,approxflow}.  
These works consider notions similar to quantitative control, while we argue that qualitative domains of control are a much better tool for bug prioritization.

{Newsome et al.~\cite{quantinfl} 
use quantitative information flow  methods to measure quantitative control. 
Their algorithm internals share similarities with S\&S, even though their goal is purely quantitative and they do not identify the key notion of domains of control. 
Additionally, their algorithm has not been designed nor used for bug prioritization, and our experiments show that S\&S performs much better for our needs.

}

\section{Conclusion}

We focused on the problem of precise and efficient bug prioritization, with the expressed goal of distinguishing more or less security-critical bugs.  
Our work on the evaluation of attacker control over vulnerability parameters to distinguish between vulnerabilities constitutes a step in this direction, yielding a theoretical framework and efficient analysis methods. 
In summary, we argue that attacker control analysis should focus on domains of control, as it allows to account for finer-grained threat models. 
Our "Shrink and Split" algorithm yields said domains of control in a scalable and flexible manner with strong formal guarantees, lending itself to practical use as shown in our experiments on real-world programs.
Future efforts could focus on applying our approach to a wider array of vulnerabilities. 

\section*{Ethical Considerations}

All vulnerabilities discussed in this work are known and/or  patched in the corresponding software's current version.  
On the other hand, if efficient and precise bug prioritization would be very beneficial to bug-fixing efforts by developers, it could also be used by malicious actors to find more promising bugs to exploit.
Nevertheless, we argue that our approach constitutes an improvement in that regard over automated exploit generation, as it does not directly enable low-skill attackers to wield ready-made exploits. 

\section*{Acknowledgements}

This work was partially supported by the “France 2030” government investment plan managed by the French National Research Agency, under the references ANR-22-PECY-0009 and ANR-22-PECY-0005.

\section*{Availability}
All our research artifacts are openly available at \url{https://doi.org/10.5281/zenodo.14699098}. 
This includes the source code of our tool Colorstreams, nix-based packaging for reproducible compilation and easy management of dependencies such as BINSEC, a docker image for easy deployment, both of our evaluation benchmarks with scripts automating the reproduction of experiments and generating figures, user tutorials and API documentation for developers.

\bibliographystyle{plain}
{\footnotesize \bibliography{bib}}

\makeatletter
\let\c@table\c@figure
\let\ftype@table\ftype@figure
\makeatother
\setcounter{figure}{2}
\renewcommand\thefigure{\Roman{figure}}  
\renewcommand\thetable{\Roman{table}}    
\renewcommand{\figurename}{Appendix}
\renewcommand{\tablename}{Appendix}
\captionsetup{labelsep=dot, labelfont=bf}

\begin{figure*}[!b]
    \caption{Domains of control for the ground-truth benchmark B1 CVEs (single-bytes excluded)}
    \label{fig:app_gt_doms}
    \includegraphics[width=\textwidth]{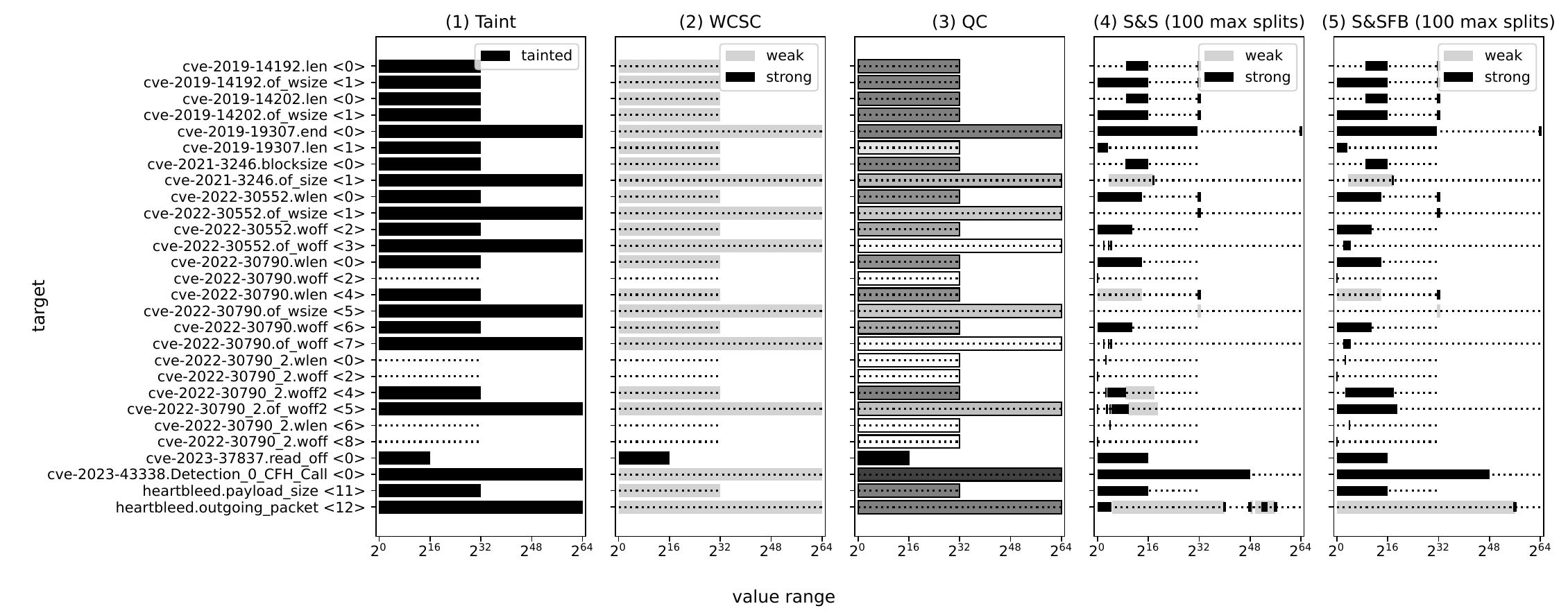}
\end{figure*}

\setcounter{figure}{0}

\begin{figure}[!t]
    \caption{Our taint propagation model}
    \label{fig:appdta}
    {
    \resizebox{\columnwidth}{!}{\begin{tabular}{l l l}
        \hline
        expression & target & taint value \\
        \hline
        \multicolumn{3}{l}{basic rules (lattice: $\bot$ --- $\top$)}\\
        \hline
        $v$ $=$ $op$ $e$ & $v$ & $t(e)$ \\
        $v = e_1$ $op$ $e_2$ & $v$ & $t(e_1) \sqcup t(e_2)$\\
        $store(addr, e)$ & $mem[addr]$ & $t(e)$ \\
        $v = load(addr)$ & $v$ & $t(mem[addr])$ \\
        $source(v)$ & $v$ & $\top$ \\
        \hline
        \multicolumn{3}{l}{(\textit{option}) propagating control-flow dependencies*} \\
        \hline
        $if(cond)$ $do$ $v = e$ & $v$ & $t(cond) \sqcup t(e)$\\
        \hline
        \multicolumn{3}{l}{(\textit{option}) over-approx memory operations when $t(addr) = \top $} \\
        \hline
        $store(addr, e)$ & $mem$ & $t(e)$ \\
        $v = load(addr)$ & $v$ & $\sqcup_{addr} t(mem[addr])$ \\
        \hline
        \multicolumn{3}{l}{(\textit{option}) local taint suppression rules (non exhaustive)} \\
        \hline
        $v = e$ $\times$ $0$ & $v$ & $\bot$ \\
        $v = e$ $-$ $e$ & $v$ & $\bot$ \\
        $if(v = const)$ & $v$ & $\bot$ \\
        \hline
    \end{tabular}}

    \par\smallskip\scriptsize{
    *Here we present a very straightforward version of control-flow dependency propagation -- more precise and practical approaches such as DTA++ only propagate taint in a few selected relevant cases \cite{dta++}.
    }\par}\medskip    

    {Variable and expressions are either untainted ($\bot$) or tainted ($\top$). $\sqcup$ is the merge operator over taint values ($\top \sqcup \bot = \top$). 
    Taint is introduced at $sources$, such as program inputs.
    Memory is represented as an array named $mem$. 
    Optionally, one may propagate taint from jump conditions to subsequent assignments or from memory addresses to written or read data in memory operations.
    Taint can also be suppressed when a single instruction reduces a tainted expression to a single value.
    The lack of constraint computation and the per-instruction granularity does not allow to propagate taint based on multi-instruction behaviours.}

    \vfill\null
\end{figure}

\begin{table}[!t]
    \caption{Detailed ground-truth benchmark B1}
    \label{tab:app_gt_bench}

        \resizebox{\columnwidth}{!}{
        \begin{tabular}{llllrHr}
\hline
 Program      & Vulnerability & Targets & Control & \multicolumn{3}{l}{Executed Instructions}\\
              & Type(s) & & Problem & Symbolic   &   Concrete &    Total \\
 \hline
 \multicolumn{4}{l}{Toy examples from Newsome et al. \cite{quantinfl}}\\
 \hline
 copy           & - & V & D &          2 &         20 &       22 \\
 ccopy          & - & V & D &          6 &         21 &       27 \\
 mcopy          & - & V & D &          3 &         20 &       23 \\
 mul            & - & V & D &          3 &         20 &       23 \\
 div            & - & V & D &          3 &         20 &       23 \\
 impflow        & - & V & D &          3 &         37 &       40 \\
 mixdup         & - & V & D &         13 &         20 &       33 \\
 popcnt         & - & V & D &         39 &         20 &       59 \\
 \hline
 \multicolumn{4}{l}{Other toy examples}\\
 \hline
 sum            & - & V & D &          4 &         28 &       32 \\
 sub            & - & V & D &          3 &         21 &       24 \\
 motex1 (vuln. a) & IUF, OOBW & S & MRW &         18 &         60 &       78 \\
 motex2 (vuln. b) & UBI, OOBW & S & MRW &          15 &       1833 &     1848 \\
 koobe (Chen et al. \cite{koobe}) & TC, PC & F & D &         12 &       4063 &     4075 \\
 uafubi         & UAF, UBI, PC & F & D &         11 &       2840 &     2851 \\
 uafubi2        & UAF, UBI, PC & F & D &          2 &       2841 &     2843 \\
 spray          & UBI & D & F &          8 &       2216 &     2224 \\
 spray2         & UBI & D & F &         16 &       2272 &     2248 \\
 spray3         & UBI & D & F &         20 &         69 &       89 \\
 can            & OOBW, PC & V, F & D &          2 &       2707 &     2709 \\
 can2           & OOBW, PC & V, F & D &          3 &       2827 &     2830 \\
 grub ($\sim$ cve-2015-8370)          & OOBW & V & D &         84 &       3784 &     3868 \\
 cfi            & - & F & D &         12 &       2237 &     2249 \\
 cfi2           & - & F & D &          5 &       2213 &     2218 \\
 minesweeper1   & OOBW, PC & R & D &        278 &      63625 &    63903 \\
 minesweeper2   & OOBW & I & MRW &        254 &     103061 &   103315 \\
 \hline
 \multicolumn{4}{l}{Real vulerabilities}\\
 \hline
 cve-2023-37837 (libjpeg) & OOBR & I & MRR &         56 &     261456 &   261512 \\
 cve-2021-3246 (libsndfile) & OOBW & S & MRW &        432 &      46440 &    46872 \\
 cve-2019-19307 (mongoose) & IOF & V & D &        104 &      67241 &    67345 \\
 cve-2019-14192 (u-boot)   & IUF, OOBW & S & MRW &         38 &       4258 &     4296 \\
 cve-2019-14202 (u-boot)   & OOBW & S & MRW &         38 &       4258 &     4296 \\
 cve-2022-30790 (u-boot) & OOBW & I, S & MRW &        161 &       3304 &     3465 \\
 cve-2022-30790-2 (u-boot) & OOBW & I & MRW &         34 &       3139 &     3173 \\
 cve-2022-30552 (u-boot) & IUF, OOBW & I, S & MRW &         91 &       1441 &     1532 \\
 heartbleed (openssl) & OOBR & S & MRR &         33 &    & 165388032 \\
 cve-2021-26567 (faad2) & OOBW, PC & R & D & 14 & & 12799 \\
 cve-2020-14393 (perl-dbi) & OOBW, PC & R & D & 2491 & & 59435770 \\
 cve-2024-41881 (sdop) & OOBW, PC & R & D & 60 & & 657357 \\
 cve-2024-43700 (xfpt) & OOBW, PC & R & D & 271 & & 179369 \\
 cve-2023-43338 (mjs) & PC & F & D & 99 & & 51401 \\
\hline
        \end{tabular}
        }
    
    \par\smallskip\scriptsize{\textbf{vulnerability types:} UAF $=$ use-after-free, UBI $=$ use-before-initialization, OOBR / OOBW $=$ out-of-bounds read / write, TC $=$ type confusion, PC $=$ pointer corruption, IUF / IOF $=$ integer underflow / overflow ---
    \textbf{targets:} R $=$ return address, F $=$ function pointer, V $=$ stack variable, I $=$ index / offset, S $=$ size ---
    \textbf{control problems:} D $=$ data, MRR / MRW $=$ memory range read / write
    }
\end{table}

\setcounter{figure}{3}

\clearpage

\begin{table}[!htb]
    \caption{Detailed realistic benchmark B2}
    \label{tab:app_real_bench}
    \resizebox{\columnwidth}{!}
    {
\begin{tabular}{lHllrrHH}
\hline
 Bug       & Analysis     & Vulnerability & Mapping & \multicolumn{2}{l}{Executed Instructions} &    SE Runtime &   Total Runtime \\
           &              & Type(s)       &         & Symbolic & Total & & \\
\hline
 PDF003    & Symbolic <6> & OOBR          & stack   &               0 &        5099250 &   0.262432    &       2479.43   \\
 PDF004    & Symbolic <5> & OOBR          & other   &            6671 &        7371051 & 406.611       &       2241.36   \\
 PDF010    & Symbolic <5> & OOBR          & other   &               0 &       14997856 &   0.59954     &       1276.8    \\
 PDF018    & Symbolic <5> & OOBR          & other   &               0 &       13291270 &   0.567884    &       1305.64   \\
 PDF019    & Symbolic <5> & OOBR          & heap    &            1463 &        7069344 &  11.336       &       2294.87   \\
 PHP011    & Symbolic <5> & OOBR          & heap    &             196 &        6059203 &   3.47398     &        380.633  \\
 PNG007    & Symbolic <5> & OOBR          & other   &               0 &          20971 &   0.000754595 &         12.4976 \\
 SQL018    & Symbolic <5> & OOBW          & heap    &               0 &         251339 &   0.0100644   &         48.976  \\
 SSL001    & Symbolic <5> & OOBR, OOBW    & heap    &            5019 &        1448773 & 251.2         &        631.029  \\
 SSL002    & Symbolic <5> & UAFW          & heap    &             295 &       10384861 &   8.56723     &       1455.74   \\
 SSL009    & Symbolic <5> & OOBR          & heap    &            3580 &        6351194 &  76.4857      &        798.597  \\
 SSLNEW001 & Symbolic <5> & UAFW          & heap    &              91 &       10408709 &   5.62858     &        802.532  \\
 SSLNEW002 & Symbolic <5> & UAFW          & heap    &             104 &       12883298 &   7.64569     &       1225.13   \\
 SSLNEW003 & Symbolic <5> & UAFW          & heap    &             284 &       10374848 &   6.94031     &       1037.58   \\
 SSLNEW004 & Symbolic <5> & UAFW          & heap    &             105 &       29008645 &  14.3769      &       2134.13   \\
 SSLNEW005 & Symbolic <5> & UAFW          & heap    &             105 &       29024109 &  13.4324      &       2008.89   \\
 SSLNEW006 & Symbolic <5> & UAFW          & heap    &             105 &       29133273 &  14.4576      &       2170.23   \\
 TIF001    & Symbolic <5> & OOBR, OOBW    & heap    &            1086 &         218407 &  77.5608      &        345.863  \\
 TIF002    & Symbolic <5> & OOBR, OOBW    & heap    &             624 &         981495 &  10.4177      &       1821      \\
 TIF002\_2 & Symbolic <5> & OOBR, OOBW    & heap    &             274 &         960672 &   1.61238     &        460.924  \\
 TIF008    & Symbolic <5> & OOBR, OOBW    & heap    &             508 &         169138 &   2.61935     &         36.6741 \\
 TIF008\_2 & Symbolic <5> & OOBR, OOBW    & heap    &             734 &         238135 &   8.64044     &        118.709  \\
 XML001    & Symbolic <6> & OOBW          & stack   &              417 &         612662 &  11.4302      &        538.431  \\
 XML002    & Symbolic <5> & OOBW          & heap    &               0 &         147239 &   0.00675035  &         49.7202 \\
 XML006    & Symbolic <6> & OOBW          & stack   &               0 &         639552 &   0.0270205   &        201.283  \\
 XML012    & Symbolic <5> & UAFR          & heap    &               0 &        7019269 &   0.29168     &        509.33   \\
\hline
\end{tabular}

    }
    {
    \par\scriptsize
    \textbf{vulnerability types:} UAFR / UAFW $=$ use-after-free read / write, OOBR / OOBW $=$ out-of-bounds read / write
    \par
    }
\end{table}

\begin{table}[!b]
    \caption{Results of Shrink and Split with different split limits on B1$+$B2}
    \label{tab:appsumsplitres}
    \resizebox{\columnwidth}{!}{
\begin{tabular}{lrlccHHHr}
\hline
 Algorithm  & Splits     & Notion &  Exact          &  Approx. ($<\times 2$)           &  False          &   TO       &   Unknown & Runtime\\
\hline
 S\&S       & 10          & DoC    &      235       &          11 (6)           &        0        &         0 &         5 & 10m56s\\
 S\&S       & 50          & DoC    &      237       &          9 (4)           &        0        &         0 &         5 & 11m12s\\
 S\&S       & 100         & DoC    &      237       &          9 (4)           &        0        &         0 &         5 & 11m52s\\
 S\&S       & 500         & DoC    &      237       &          9 (4)           &        0        &         0 &         5 & 15m20s\\
 S\&S       & 1000        & DoC    &      237       &          9 (4)           &        0        &         0 &         5 & 19m33s\\
 \hline
 S\&SFB     & 10          & DoC    &      238       &          8 (7)           &        0        &         0 &         5 & 10m26s\\
 S\&SFB     & 50          & DoC    &      240       &          6 (5)           &        0        &         0 &         5 & 10m45s\\
 S\&SFB     & 100         & DoC    &      241       &          5 (4)           &        0        &         0 &         5 & 10m55s\\
 S\&SFB     & 500         & DoC    &      241       &          5 (4)           &        0        &         0 &         5 & 13m05s\\
 S\&SFB     & 1000        & DoC    &      241       &          5 (4)           &        0        &         0 &         5 & 15m29s\\
\hline
\end{tabular}
    }
    \smallskip\par
    Increasing the split limit for Shrink and Split has a limited impact on precision on our benchmark.
    Since holes are likely to occur at regular intervals in cases where they are numerous, other solutions such as taking fixed bits into account improve precision more effectively.
    Higher split limits only affect runtimes marginally.
\end{table}

\begin{figure}[!t]
    \caption{Cactus plot of all tested algorithms on B1$+$B2}
    \label{fig:cactus}
    \includegraphics[width=\columnwidth]{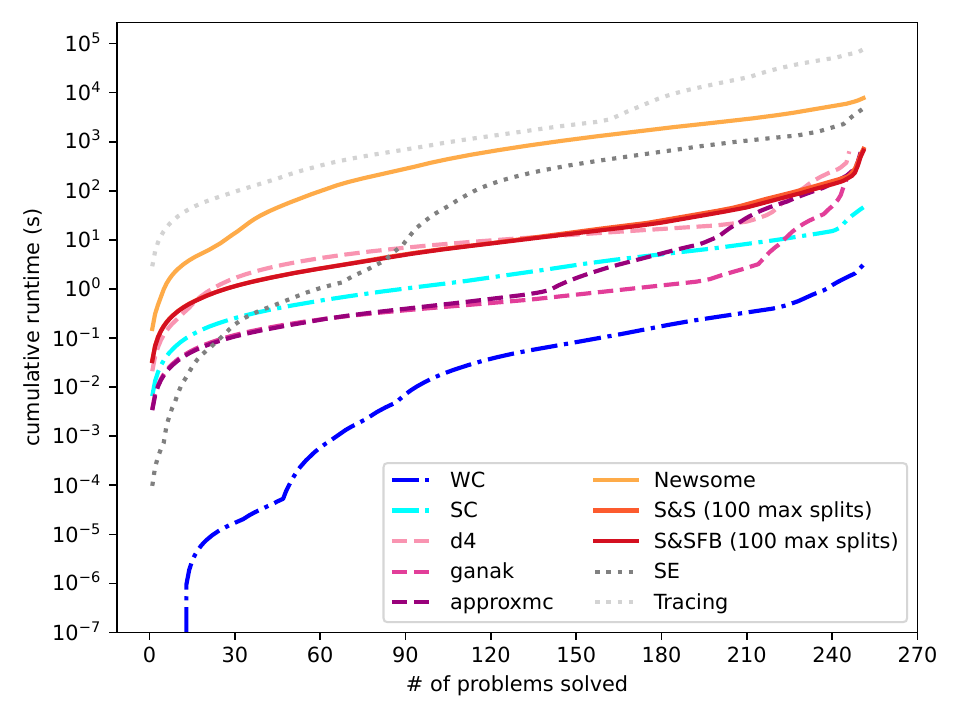}
    {
    Tracing and symbolic execution are common to all other algorithms. 
    The sum of both indicates the time needed to derive path constraints.
    
    \par
    }
\end{figure}

\begin{table}[!b]
    \caption{Testing multiple weight functions for scoring the OOB vulnerabilities from Table \ref{tab:scores}}
    \label{tab:appwfunscores}
    \resizebox{\columnwidth}{!}{
    \begin{tabular}{l H H l c l c l c}
        \hline
        Vulnerability & & & \multicolumn{2}{l}{$\omega: x \mapsto \frac{1}{ln(2)x}$} & \multicolumn{2}{l}{$\omega': x \mapsto \frac{1}{x^2}$} & \multicolumn{2}{l}{$\omega'': x \mapsto \frac{1}{\sqrt{x}}$} \\
        \hline
        \multicolumn{9}{c}{OOB writes} \\
        \hline
        motex1 & \textit{fixed} & of\_wsize & 0 & \Smiley[1.5][green] & 0 & \Smiley[1.5][green] & 0 & \Smiley[1.5][green] \\
        motex2 & \textit{fixed} & of\_wsize & 5.36 & \Neutrey[1.5][yellow] & 15.23 & \Sadey[1.5][red!50!white] & 7.9e-8 & \Neutrey[1.5][yellow]\\
        cve-2021-3246 & \textit{fixed} & of\_size & 14.14 & \Sadey[1.5][red!50!white] & 5.33 & \Neutrey[1.5][yellow] & 6.5e-6 & \Neutrey[1.5][yellow]\\
        cve-2019-14192 & \textit{fixed} & of\_wsize & 15.97 & \Sadey[1.5][red!50!white] & 16 & \Sadey[1.5][red!50!white] & 0.12 & \Sadey[1.5][red!50!white] \\
        cve-2019-14202 & \textit{fixed} & of\_wsize & 15.97 & \Sadey[1.5][red!50!white] & 16 & \Sadey[1.5][red!50!white] & 0.12 & \Sadey[1.5][red!50!white] \\
        cve-2022-30790 & of\_woff(\_2) & of\_wsize(\_2) & 0.51 & \Smiley[1.5][green] & 1.75 & \Smiley[1.5][green] & 8.4e-9 & \Smiley[1.5][green] \\
        cve-2022-30552 & of\_woff & of\_wsize & 0.51 & \Smiley[1.5][green] & 1.75 & \Smiley[1.5][green] & 8.4e-9 & \Smiley[1.5][green]\\
        cve-2022-30790-2 & of\_woff2 & \textit{fixed} & 2.37 & \Neutrey[1.5][yellow] & 8 & \Neutrey[1.5][yellow] & 1.3e-6 & \Neutrey[1.5][yellow]\\
        \hline
        \multicolumn{9}{c}{OOB reads}\\
        \hline
        cve-2023-37837 & read\_off & \textit{fixed} & 16 & \Sadey[1.5][red!50!white] & 16 & \Sadey[1.5][red!50!white] & 16 & \Sadey[1.5][red!50!white] \\
        heartbleed & \textit{fixed} & payload\_size & 16 & \Sadey[1.5][red!50!white] & 32 & \Sadey[1.5][red!50!white] & 0.12 & \Sadey[1.5][red!50!white] \\
        \hline
    \end{tabular}
    }
    {
        \par\smallskip
        Compared to $\omega$, $\omega'$ favors smaller values more while $\omega''$ is less biased toward them.
        This results in some differences, especially regarding \textit{motex2} and \textit{cve-2021-3246}.
        However the general tendency remains similar overall.
    }

\end{table} 

\FloatBarrier

\section*{Appendix VIII. Manual Evaluation of \textit{cve-2022-30790} and \textit{cve-2022-30552}}

\label{app:ubootdefrag}

{Both vulnerabilities occur in uboot v2022.01, in the \textit{\_\_net\_defragment} function from \textit{net/net.c}, line 900 \cite{ubootdefrag}.
This function takes packets containing data fragments and puts them back together into a static buffer.
A linked list structure is used to keep track of holes in the data and check if the incoming fragments fit into them.}

\begin{lstlisting}[
        language=c, 
        caption=Vulnerable \_\_net\_defragment function from uboot v2022.01, 
        label=lst:defrag,
        basicstyle=\scriptsize,
        commentstyle=\color{gray},
        numbers=left,
        xleftmargin=3em,
        xrightmargin=1em,
        frame=single,
        captionpos=b
    ]
static struct ip_udp_hdr *__net_defragment
    (struct ip_udp_hdr *ip, int *lenp)
{
    /*static buffer where the data is assembled*/
    static uchar pkt_buff[IP_PKTSIZE];
    ...
    /*hole linked list metadata struct*/
    struct hole *payload, *thisfrag, *h, *newh;
    ...
    uchar *indata = (uchar *)ip;
    int offset8, start, len, done = 0;
    /*data fragment offset from packet header*/
    u16 ip_off = ntohs(ip->ip_off);

    /*start of data in pkt_buff*/
    payload = (struct hole *)(pkt_buff 
        + IP_HDR_SIZE);
    offset8 =  (ip_off & IP_OFFS);
    /*start of incoming fragment in pkt_buff*/
    thisfrag = payload + offset8;
    start = offset8 * 8;
    /*data fragment length computation*/
    /*can go negative (cve-2022-30552)*/
    len = ntohs(ip->ip_len) - IP_HDR_SIZE;

    /*Here the program checks if the data fragment 
    would overflow pkt_buff. However a negative 
    len can pass if start >= IP_HDR_SIZE.*/
    if (start + len > IP_MAXUDP)
	return NULL;

    ...
    /*Here the function checks if the packet is 
    coherent with already received fragments, i.e., 
    if it fits into a hole in pkt_buff. These holes 
    are tracked via metadata stored into them akin 
    to a linked list. If there is a conflict, the 
    function aborts.*/
    /*The function also updates the metadata, 
    which can be corrupted (cve-2022-30790).*/
    /*thisfrag and len are NOT updated.*/
    ...

    /*thisfrag and len are independent from the 
    hole metadata, which is only used to auth-
    orize the write. Therefore an overflow can
    only happen if len is negative and underflows
    due to an implicit cast to size_t 
    (cve-2022-30552).*/
    memcpy((uchar *)thisfrag, indata + IP_HDR_SIZE, 
        len);
    ...
}
\end{lstlisting}

{Listing \ref{lst:defrag} gives a summary of the manual analysis of \textit{cve-2022-30552} and \textit{cve-2022-30790} and explains why \textit{cve-2022-30790} cannot lead to arbitrary writes outside of the target buffer on line $50$ as was previously thought \cite{nccgroup}.
In particular, the linked list metadata corruption cannot lead to bypassing the check on line $29$.
Our tool shows the same capabilities for both vulnerabilities due to \textit{cve-2022-30552} being triggerable within \textit{cve-2022-30790}'s execution path.}

{On the other hand, \textit{cve-2022-30790} enables some limited out-of-bounds writes during the linked list update, which were not discussed in the original human analysis.
We analyzed those capabilities as \textit{cve-2022-30790-2}.}

\section*{Appendix IX. Magma Vulnerability Scores Discussion}

\label{app:rq4interp}

\setcounter{figure}{5}
\renewcommand\thetable{\arabic{figure}}
\renewcommand{\tablename}{Table}
\captionsetup{labelsep=colon, labelfont=normalfont}

Our results for the Magma vulnerabilities from RQ4 (Section \ref{sec:rq4}) illustrate the benefits of our approach as well as areas of possible improvements.

\begin{table}[!htb]
    \caption{Detailed scores for the magma vulnerabilities (B2)}
    \label{tab:detailedscoresmagma}
    \resizebox{\columnwidth}{!}{
\begin{tabular}{llrHllll}
\hline
 Bug       &   CVE & CVSS   & Mapping   &\multicolumn{4}{l}{wQC Score} \\
           &      &        &           &   Base  &   Size  &   Data  &   Overall \\
\hline
\multicolumn{8}{c}{OOB writes}\\
\hline
 SSLNEW001 &    - & -  & heap      &     0.00477  &     0.703    &      1       &         0.569   \\
 SSLNEW003 &    - & -  & heap      &     0.00477  &     0.703    &      1       &         0.569   \\
 SSL002    &    CVE-2016-6309 &  9.8  & heap      &     0.00477  &     0.703    &      1       &         0.569   \\
 SSLNEW005 &    - & -  & heap      &     0.00473  &     0.697    &      1       &         0.567   \\
 SSLNEW006 &    - & -  & heap      &     0.00473  &     0.697    &      1       &         0.567   \\
 SSLNEW004 &    - & -  & heap      &     0.00473  &     0.697    &      1       &         0.567   \\
 SSLNEW002 &    - & -  & heap      &     0.00469  &     0.696    &      1       &         0.567   \\
 TIF002\_2 &    CVE-2016-5314 & 8.8  & heap      &     0.022    &     0.643    &      1       &         0.555   \\
 TIF002    &    CVE-2016-5314 & 8.8  & heap      &     0.0223   &     0.00512  &      1       &         0.342   \\
 \arrayrulecolor{gray!50!white}\hline\arrayrulecolor{black}
 TIF008    &    CVE-2015-8784 & 6.5  & heap      &     0.0285   &     0.0587   &      0.00391 &         0.0304  \\
 TIF008\_2 &    CVE-2015-8784 & 6.5  & heap      &     0.02     &     0.0587   &      0.00391 &         0.0275  \\
 SSL001    &    CVE-2016-2108 & 9.8  & heap      &     0.0208   &     0.0554   &      0.00391 &         0.0267  \\
 TIF001    &    CVE-2016-9535 & 9.8  & heap      &     0.0192   &     0.0554   &      0.00391 &         0.0262  \\
 XML006    &    CVE-2017-9048 & 7.5  & stack     &     0.0173   &     0.0342   &      0.00391 &         0.0185  \\
 XML001    &    CVE-2017-9047 & 7.5  & stack     &     3.27e-05 &     6.76e-05 &      0.0508  &         0.017   \\
 XML002    &    CVE-2017-0663 & 7.8  & heap      &     0.000912 &     0.0178   &      0.00391 &         0.00755 \\
 SQL018    &    CVE-2015-3414 & 7.5  & heap      &     0.00398  &     0.0102   &      0.00391 &         0.00602 \\
 \hline
 \multicolumn{8}{c}{OOB reads}\\
 \hline
 PDF019    &    CVE-2017-9776 & 7.8  & heap      &     0.699    &     0.0535   &      0       &         0.376   \\
 PHP011    &    CVE-2018-14883 & 7.5  & heap      &     0.000181 &     0.447    &      0       &         0.224   \\
 \arrayrulecolor{gray!50!white}\hline\arrayrulecolor{black}
 PDF003    &    CVE-2017-9865 & 5.5  & stack     &     0.0303   &     0.0585   &      0       &         0.0444  \\
 TIF008    &    CVE-2015-8784 & 6.5*  & heap      &     0.0285   &     0.0587   &      0       &         0.0436  \\
 SSL001    &    CVE-2016-2108 & 9.8*  & heap      &     0.027    &     0.0554   &      0       &         0.0412  \\
 TIF008\_2 &    CVE-2015-8784 & 6.5*  & heap      &     0.02     &     0.0587   &      0       &         0.0393  \\
 TIF001    &    CVE-2016-9535 & 9.8*  & heap      &     0.0192   &     0.0554   &      0       &         0.0373  \\
 SSL009    &    CVE-2017-3735 & 5.3  & heap      &     0.0287   &     0.0313   &      0       &         0.03    \\
 XML012    &    CVE-2016-1836 & 5.5  & heap      &     0.000277 &     0.0537   &      0       &         0.027   \\
 TIF002    &    CVE-2016-5314 & 8.8*  & heap      &     0.0223   &     0.00512  &      0       &         0.0137  \\
 TIF002\_2 &    CVE-2016-5314 & 8.8*  & heap      &     0.022    &     0.00104  &      0       &         0.0115  \\
 PDF004    &    CVE-2019-10873 & 6.5  & other     &     0        &     0        &      0       &         0       \\
 PDF010    &    - & -  & other     &     0        &     0        &      0       &         0       \\
 PDF018    &    CVE-2018-10768 & 6.5  & other     &     0        &     0        &      0       &         0       \\
 PNG007    &    CVE-2013-6954 & 5  & other     &     0        &     0        &      0       &         0       \\
\hline
\end{tabular}
    }
    {
        \par \smallskip \scriptsize
        *likely based on OOB writes\\
        grey lines represent score cutoff points
        \par
    }
\end{table}

\myparagraph{Contrast between different vulnerability capabilities}
At a glance, our results from Figure \ref{fig:magma_scores} and Table \ref{tab:detailedscoresmagma} show a clear difference in OOB capabilities between the vulnerabilities.
While it is not feasible to analyze each vulnerability in depth, we can still check that the scores are appropriate by manually reviewing and interpreting the domains of control and other readily available information.

On the side of OOB writes, we can see a clear discrepancy between highly controllable and uncontrollable vulnerabilities. 
In particular, \textit{SSL002} and the \textit{SSLNEW} vulnerabilities grant attackers the ability to write arbitrary data over a controlled amount of bytes, starting at a static address.
This alone makes these vulnerabilities quite dangerous and grants them a high score, but furthermore, we know that they are heap use-after-frees, as shown in Appendix \ref{tab:app_real_bench}. 
Hence the base address of the write could be indirectly influenced via heap layout manipulation.

In comparison, \textit{TIF002} only grants control over the written data, which could still be useful to an attacker, hence its middling score.
However, other vulnerabilities such as \textit{TIF001} cannot be controlled much, if at all, and have a low score as a result.
They could only be exploited if, by luck, they align perfectly with exploit requirements.
\textit{TIF001} in particular is an \textit{off-by-one} OOB write which does not allow control over the written data.

Regarding OOB reads, \textit{PDF019} and \textit{PHP011} offer the widest capabilities, although this is arguably less important for these vulnerabilities to be exploitable, since useful information can be extracted from a large read and overwriting critical data in a detrimental way is not a concern.
Still, our results show that \textit{PDF004}, \textit{PDF010}, \textit{PDF018} and \textit{PNG007} cannot be valid writes and thus can only be used to cause crashes.
Their score is, appropriately, zero.

\myparagraph{Correlation with CVSS scores}
While CVSS scores are worst-case and qualitative at best, it is interesting to note that we still observe a light correlation between them and our scores, especially for OOB writes.
For those, we only find two outliers, \textit{SSL001} and \textit{TIF001}, with high CVSS scores but low wQC scores.
Both analyzed writes are \textit{off-by-one} without controlled data, and thus would be difficult to exploit.
However, note that in both cases we only included results for the first write out of several, which all exhibit roughly the same characteristics and are not contiguous as could be expected from a loop.
Combining all information could results in higher capabilities, however the lack of control on the written data is a hard limit on exploitability.

For OOB reads, some vulnerability also cause OOB writes, which explains their higher scores.
It is also interesting to note that \textit{PDF010} grants the same capabilities as \textit{PDF004}, \textit{PDF018} and \textit{PNG007}, the ability to cause crashes with an invalid read, yet does not have an associated CVE.

\myparagraph{Analysis of new vulnerabilities}
The \textit{SSLNEW} vulnerabilities are not technically part of the magma benchmark.
They were triggered during the original Magma evaluation but not caught by any of Magma's bug oracles.
From our results, we can see that they have very similar capabilities to \textit{SSL002}, as well as the same type, namely \textit{heap use-after-free}.
Combined with the detection reports from ASAN, this suggests that these vulnerabilities are impacts of a same bug, though through different execution contexts.
This intuition is validated by the fact that all these vulnerabilities are fixed by the same patch.

\myparagraph{Illustration of limitations}
In some cases, it is possible that vulnerabilities have greater capabilities than we were able to measure.
For example, the size of the OOB write in \textit{XML001} could be controlled implicitly via data as this vulnerability occurs during the parsing of a file, which mainly consists in string operations.
Additionally, the repeated reads in \textit{XML012} suggest that they could occur within a loop, of which the bounding condition could be controlled.
Alas, detecting these parameters and tracking their constraints remains challenging for binary-level analysis, as discussed in Section \ref{sec:disc}.

\section*{Appendix X. Evaluation of Weak Interval Densities in Domains computed with Shrink and Split}

\label{tab:snscount}

{In an ideal scenario, Shrink and Split yields exact domains of control.
However, in practice, analysis may be interrupted or proving strong control on an interval may fail.
This leads to only weak control being guaranteed on some intervals, while their actual density may change the interpretation of results.}

{To evaluate the impact of weak intervals on result interpretation, we computed the densities of those occurring in our analysis of benchmarks B1 and B2. 
We used ganak to count feasible values, then divided the result by the width of the interval to obtain its density.
We then computed the average of these densities for each case.}

\begin{table}[!htb]
    {
    \caption{Weak interval densities on benchmarks B1 and B2}
    \label{tab:snscount_tab}
\resizebox{\columnwidth}{!}{\begin{tabular}{lHHHHlHHHlHc}
\hline
 Program                   & Sink         & Size   & \multicolumn{4}{l}{S\&S}                 & \multicolumn{4}{l}{S\&SFB} & Evaluation Impact*               \\
\hline
 uafubi           & foo <0>      & 64     & 0.985 & - & 0.985 & 0       & 0.987 & - & 0.987 & 1  & -     \\
 mul              & j <0>        & 32     & 0.5 & - & 0.5 & 0           & - & - & no weak & 0   & -      \\
 mixdup           & i <0>        & 32     & 1.81e-05 & - & 1.81e-05 & 0 & 1.81e-05 & - & 1.81e-05 & 0 & - \\
 grub             & canary <0>   & 32     & 5.7e-06 & - & 5.7e-06 & 0   & - & - & no weak & 0  & -       \\
 cve-2022-30790\_2$^1$ & of\_woff2 <5> & 64     & 0.125 & - & 0.125 & 0       & - & - & no weak & 0       & $=$  \\
 cve-2022-30790\_2$^2$ & woff2 <4>    & 32     & 0.25 & - & 0.25 & 0         & - & - & no weak & 0        & $=$ \\
 cve-2022-30790$^3$   & of\_wsize <5> & 64     & - & 1 & 1 & 0               & - & 1 & 1 & 0        & $=$ \\
 cve-2022-30790$^4$   & wlen <4>     & 32     & - & 1 & 1 & 0               & - & 1 & 1 & 0         & $=$\\
 cve-2021-3246    & of\_size <1>  & 64     & timeout & - & timeout & 1         & timeout & - & timeout & 1  & timeout        \\
 heartbleed$^5$ & outgoing\_packet <12> & 64 & 0.00117 & - & 0.00117 & & 6.04e-08 & - & 6.04e-08 & & $=$ \\
 SSL002$^6$           & Detection\_0\_OOB\_write\_size \ensuremath{<}1\ensuremath{>}         & 64     &  -  & 1 & 1 & 0               & - & 1 & 1 & 0      & $=$          \\
 PDF019$^7$           & Detection\_0\_OOB\_read\_base \ensuremath{<}0\ensuremath{>}          & 64     & - & - & timeout & 2         & - & - & timeout & 2   & timeout      \\

 \hline
 Average                   &              &        & 0.266 & 1 & 0.486 & 3           & 0.329 & 1 & 0.664 & 4          \\
\hline
\end{tabular}}
    {
        \par \smallskip \scriptsize
        
        * impact on the score category from Table \ref{tab:scores} (\Smiley[1][green] / \Neutrey[1][yellow] / \Sadey[1][red!50!white]). 
        $\uparrow$: up a category, $=$: no change, $\downarrow$: down a category ---
        $^1$of\_woff2 <5>, $^2$woff2 <4>, $^3$of\_wsize <5>, $^4$wlen <4>, $^5$outgoing\_packet <12> (leak), $^6$write size, $^7$read base address
        \par
    }}
\end{table}

{As shown in Table \ref{tab:snscount_tab}, the densities of weak intervals span a wide range, with some almost empty and other completely full. 
As such, a systematic way of correctly handling them in practice seems out of reach.  
Nevertheless, in our experiments, these refined results would not change our score categories, solidifying our prior findings against the manual ground truth.}

\section*{Appendix XI. Impact of Constraint Relaxation on Domains of Control}

\label{tab:app_relax}

{One way to improve the scalability of symbolic execution is to remove some constraints.
This approach is called constraint relaxation or under-constrained symbolic execution \cite{relax} and may cause over-approximation.
It may thus affect bug prioritization with our method if used.}

{To evaluate the impact of constraint relaxation on our method, we computed domains of control with S\&SFB (100 maximum splits) based on the constraints from the vulnerable function only for each real-world vulnerability in benchmark B1 (except those where the vulnerable function was already the main analyzed function).
We then computed their size ratio compared to the domains obtained in our mainline experiments.}

\begin{table}[!htb]
    {
    \caption{Domain size ratios with and without relaxation for benchmark B1 (S\&SFB, 100 max. splits)}
    \label{tab:app_relax_tab}
\resizebox{\columnwidth}{!}{\begin{tabular}{lclr}
\hline
 Bug          & Eval.   & Sink                            &   Domain  \\
              & Impact* &                                 & Ratio     \\
\hline
 heartbleed & $=$ & payload\_size & 1  \\

 cve-2024-43700 & $=$ & Detection\_0\_CFH\_Ret\_byte\_(7-0) \ensuremath{<}7-0\ensuremath{>}  &                 1.01      \\

 cve-2024-41881 & $=$ & Detection\_0\_CFH\_Ret\_byte\_(7-0) \ensuremath{<}7-0\ensuremath{>}  &                 1        \\
 cve-2021-3246 & $=$  & of\_size \ensuremath{<}1\ensuremath{>}                     &                 9.39e+13  \\
   & & blocksize \ensuremath{<}0\ensuremath{>}                   &                 3.3e+04   \\

 cve-2020-14393 & $=$ & Detection\_0\_CFH\_Jump\_byte\_(7-0) \ensuremath{<}7-0\ensuremath{>} &                 1         \\
 cve-2019-19307 & $=$ & len \ensuremath{<}1\ensuremath{>}                         &                 1      \\
   & & end \ensuremath{<}0\ensuremath{>}                         &                 1        \\
 cve-2019-14202 & $=$ & of\_wsize \ensuremath{<}1\ensuremath{>}                    &                 6.67e+04  \\
 &  & len \ensuremath{<}0\ensuremath{>}                         &                 6.61e+04   \\
 cve-2019-14192 & $=$ & of\_wsize \ensuremath{<}1\ensuremath{>}                    &                 6.67e+04 \\
 & & len \ensuremath{<}0\ensuremath{>}                         &                 6.61e+04 \\
\hline
\end{tabular}}

{\par\smallskip\scriptsize
*impact on the score category from Table \ref{tab:scores} (\Smiley[1][green] / \Neutrey[1][yellow] / \Sadey[1][red!50!white]). 
$\uparrow$ up a category, $=$ no change, $\downarrow$ down a category
\par}
}
\end{table}

{As expected, Table \ref{tab:app_relax_tab} shows that the domains of control with relaxed constraints are always equal or larger than those for the exact constraints (over-approximation). More precisely, the domains are very similar in half the cases (6 out of 12), but we observe very significant difference on the other half (6 out of 12). 
Yet, interestingly, our score categorization from Table \ref{tab:scores} would however not be affected by this approximation, as the largest ratios occur here for vulnerabilities with already high capabilities. 
In terms of performance, symbolic execution runtime ranges from unaffected to substantially reduced (28x faster), yet   
the overall analysis runtime is only marginally affected (2x faster at best), since tracing takes the most time due to the large difference between the size of the whole trace and what is symbolically followed (see Appendix \hyperref[fig:cactus]{VI}).}

\section*{Appendix XII. Impact of Input Concretization on Domains of Control}

\label{tab:app_concr}

{Another way of improving the scalability of symbolic execution is to concretize part of the symbolic inputs, in order to reduce the complexity of the constraints and the number of symbolic state updates \cite{dart, concr}.
This approach may cause under-approximation and thus affect bug prioritization.
}

{To evaluate the impact of input concretization on our method, we computed domains of control with S\&SFB (100 maximum splits) for each real-world vulnerability from benchmark B1 with {an arbitrary} part of the input concretized.
We then computed the size ratio compared to the domains obtained in our mainline experiments.}

\begin{table}[!htb]
    {
    \caption{Domain size ratios with and without partial input concretization for benchmark B1 (S\&SFB, 100 max. splits)}
    \label{tab:app_concr_tab}
\resizebox{\columnwidth}{!}{\begin{tabular}{lclr}
\hline
 Bug              & Eval. & Sink                            &   Domain  \\
                  & Impact* &                                 & Ratio  \\
\hline

 cve-2024-43700 & $=$   & Detection\_0\_CFH\_Ret\_byte\_(7-4) \ensuremath{<}7-4\ensuremath{>}  &                 0.00403   \\

    & & Detection\_0\_CFH\_Ret\_byte\_(3-0) \ensuremath{<}3-0\ensuremath{>}  &                 1         \\

 cve-2024-41881 & $=$  & Detection\_0\_CFH\_Ret\_byte\_(7-4) \ensuremath{<}7-4\ensuremath{>}  &                 0.00394   \\

 & & Detection\_0\_CFH\_Ret\_byte\_(3-0) \ensuremath{<}3-0\ensuremath{>}  &                 1         \\
 cve-2023-37837 & $=$  & read\_off \ensuremath{<}0\ensuremath{>}                    &                 0.00391   \\
 cve-2022-30790\_2 & $=$ & woff \ensuremath{<}8\ensuremath{>}                        &                 1        \\
 & & wlen \ensuremath{<}6\ensuremath{>}                        &                 1        \\
 & & of\_woff2 \ensuremath{<}5\ensuremath{>}                    &                 1       \\
 & & woff2 \ensuremath{<}4\ensuremath{>}                       &                 1        \\
 & & woff \ensuremath{<}2\ensuremath{>}                        &                 1        \\
 & & wlen \ensuremath{<}0\ensuremath{>}                        &                 1        \\
 cve-2022-30790 & $\downarrow$  & of\_woff <7> & 0** \\
 &  & woff \ensuremath{<}6\ensuremath{>}                        &                 0.995  \\
 & & of\_wsize <5> & 0**  \\
 &   & wlen \ensuremath{<}4\ensuremath{>}                        &                 6.11e-05 \\
 &   & woff \ensuremath{<}2\ensuremath{>}                        &                 1      \\
 &  & wlen \ensuremath{<}0\ensuremath{>}                        &                 6.11e-05 \\
 cve-2022-30552 & $=$ & of\_woff <3> & 0  \\
 &   & woff \ensuremath{<}2\ensuremath{>}                        &                 0.000488  \\
 &   & of\_wsize \ensuremath{<}1\ensuremath{>}                    &                 0.000489  \\
 &   & wlen \ensuremath{<}0\ensuremath{>}                        &                 0.00104   \\
 cve-2021-3246  & $\downarrow$  & of\_size \ensuremath{<}1\ensuremath{>}                     &                 0.00288   \\
 &    & blocksize \ensuremath{<}0\ensuremath{>}                   &                 0.00393  \\

 cve-2021-26567 & $=$  & Detection\_0\_CFH\_Ret\_byte\_(7-4) \ensuremath{<}7-4\ensuremath{>}  &                 0.00392   \\

 &   & Detection\_0\_CFH\_Ret\_byte\_(3-0) \ensuremath{<}3-0\ensuremath{>}  &                 1         \\
 cve-2020-14393  & $\downarrow$ & Detection\_0\_CFH\_Jump\_byte\_7 \ensuremath{<}7\ensuremath{>} &                 1       \\
 &   & Detection\_0\_CFH\_Jump\_byte\_6 \ensuremath{<}6\ensuremath{>} &                 1        \\
 &   & Detection\_0\_CFH\_Jump\_byte\_5 \ensuremath{<}5\ensuremath{>} &                 0.00781  \\
 &   & Detection\_0\_CFH\_Jump\_byte\_4 \ensuremath{<}4\ensuremath{>} &                 0.00391  \\
 &   & Detection\_0\_CFH\_Jump\_byte\_3 \ensuremath{<}3\ensuremath{>} &                 0.00391  \\
 &   & Detection\_0\_CFH\_Jump\_byte\_2 \ensuremath{<}2\ensuremath{>} &                 0.00391  \\
 &   & Detection\_0\_CFH\_Jump\_byte\_1 \ensuremath{<}1\ensuremath{>} &                 0.5      \\
 &   & Detection\_0\_CFH\_Jump\_byte\_0 \ensuremath{<}0\ensuremath{>} &                 1        \\
 cve-2019-19307 & $=$  & len \ensuremath{<}1\ensuremath{>}                         &                 0.545    \\
 &   & end \ensuremath{<}0\ensuremath{>}                         &                 3.81e-06 \\
 cve-2019-14202  & $\downarrow$ & of\_wsize \ensuremath{<}1\ensuremath{>}                    &                 1.55e-05  \\
 &  & len \ensuremath{<}0\ensuremath{>}                         &                 1.54e-05  \\
 cve-2019-14192 & $\downarrow$  & of\_wsize \ensuremath{<}1\ensuremath{>}                    &                 1.55e-05  \\
 &   & len \ensuremath{<}0\ensuremath{>}                         &                 1.54e-05  \\
\hline
\end{tabular}}

{\par\smallskip\scriptsize
*impact on the score category from Table \ref{tab:scores} (\Smiley[1][green] / \Neutrey[1][yellow] / \Sadey[1][red!50!white]). 
$\uparrow$: up a category, $=$: no change, $\downarrow$: down a category --- **here the feasible values do not even allow OOBs\par}
}
\end{table}

{As shown in Table \ref{tab:app_concr_tab}, and as expected, the resulting partially concretized domains of control are always equal or smaller than the exact ones (under-approximation).  
More precisely, the domains are very similar ($>0.99$) in 14 out of 39 cases and significantly different ($<0.1$) in 22. For 5 out of 12 vulnerabilities, our assessment from Table \ref{tab:scores} would have been affected, with lower scores leading to a lower threat classification.
Similarly to relaxation, symbolic execution can be faster (up to 83x), although overall performance gains are again small due to the tracing runtime.
}

\end{document}